\documentclass[aps,preprint,superscriptaddress,groupedaddress]{revtex4}  
\usepackage{graphicx}  
\usepackage{dcolumn}   
\usepackage{bm}        
\usepackage{amssymb}   
\usepackage{amsmath}
\usepackage{latexsym}
\usepackage{verbatim}
\usepackage{amsfonts}
\usepackage{color}
\usepackage{subfigure}

\begin{document}


\title{Group percolation in interdependent networks}

\author{Zexun Wang}\thanks{These two authors contributed equally}
\affiliation{School of Data and Computer Science, Sun Yat-sen University, Guangzhou 510006, China}
\affiliation{Big Data Research Center,University of Electronic Science and Technology of China, Chengdu 611731, China}
\author{Dong Zhou}\thanks{These two authors contributed equally}
\affiliation{Simula Research Laboratory, 1325 Lysaker, Norway}
\author{Yanqing Hu}\email{yanqing.hu.sc@gmail.com}
\affiliation{School of Data and Computer Science, Sun Yat-sen University, Guangzhou 510006, China}
\affiliation{Big Data Research Center, University of Electronic Science and Technology of China, Chengdu 611731, China}

\date{\today}

\begin{abstract}
In many real network systems, nodes usually cooperate with each other and form groups, in order to enhance their robustness to risks. This motivates us to study a new type of percolation, group percolation, in interdependent networks under attacks. In this model, nodes belonging to the same group survive or fail together. We develop a theoretical framework for this novel group percolation and find that the formation of groups can improve the resilience of interdependent networks significantly. However, the percolation transition is always of first order, regardless of the distribution of group sizes. As an application, we map the interdependent networks with inter-similarity structures, which attract many attentions very recently, onto the group percolation and confirm the non-existence of continuous phase transitions.

\end{abstract}
\maketitle

\section{Introduction}
Complex networks theory provides powerful tools for modelling topological properties of complex systems. Percolation of complex networks concentrates on the relation between node or link attacks/failures and the remaining functional largest component (giant component), which describes the robustness of different sorts of single networks under attacks/failures \cite{cohen2010complex,newman2010networks,NewmanPRL2000,NewmanPRE2001,Shlomo2000,Mendes2001PRE,CohenPREpcScalefree2002,CohenDirectedPRE2002,Schneider2011}. However, most real systems are not isolated, but coupled. Due to the existence of inter-dependency relationships, failures in one network may propagate across different systems, and lead to larger-scale failures to the whole system. In 2010, the percolation theory of interdependent networks has been introduced, presenting abrupt percolation transitions where the giant component suddenly disappears \cite{buldyrev2010catastrophic}. After that, interdependent networks has been extensively investigated after various generalizations \cite{parshani2010interdependent,Parshani2011,Hu2011percolation,gao2012networks,li2012cascading,baxter2012avalanche,hu2013percolation,bashan2013extreme,cellai6359percolation,ZhouDongPRE2014,YanqingNP2014,BianconiPRE2014,GohPRECorr2014,feng2015simplified,Cellai2016PRE,Baxter2016PRE,Hackett2016PRX,Kleineberg2017PRL,Yuan2017PNAS}.

Until now, most studies on interdependent networks still consider interdependency relations between single nodes. However, under certain circumstances some nodes within one network tend to fail or survive together.
One example is the percolation of interdependent networks with inter-similarity~\cite{hu2013percolation}. While common links (links shared by subnetworks) are introduced into a system of two interdependent networks, it has been shown that all nodes that are connected via paths of common links will fail or survive together, since they must belong to the mutual giant component or not simultaneously~\cite{hu2013percolation}. In the above mentioned specific work, the approach of combining groups of nodes (i.e., components of common links) into ``super-nodes'' (each super-node corresponds to a group of nodes) has been applied, where links are also merged on a super-node if they have at least one end belonging to the corresponding node group~\cite{hu2013percolation}. However, systematic explorations of a generalized model of interdependent networks with combined node groups are still lacking.

In fact, percolation of interdependent networks with combined node groups is of great importance in both analytical and practical aspects. On one hand, it presents solid analytical tools for solving the percolation models where nodes fail/survive in groups, as in \cite{hu2013percolation}. On the other hand, combining a group of regular nodes can cause a larger degree value of the constructed super-node, since all links related to these regular nodes will be merged on the combined super-node. Therefore, this model proposes a new approach for modelling percolation/dependency failures where nodes tend to mitigate risks by establishing collaborations and behave together as groups. Notice that this is different with previous percolation studies where more nodes fail together due to dependency links within single networks \cite{Parshani2011,Liu2016}. Examples of real world systems where node groups are formed to increase robustness include companies that reduce risks through cooperation strategies. In the mobile Internet, an individual mobile device sometimes shares its Internet connections with others if they have lost the connections. In this scenario, different devices also tend to keep the Internet connections or not (fail) together, as a group. Therefore, it is still an important open question how these node grouping phenomena in real world systems impact the robustness of the entire systems. 

In this work, we for the first time develop a general model of ``group percolation'' in interdependent networks, which provides a new concept in complex network studies. In this model, regular nodes are randomly divided into node groups, and each node group is contracted into a ``super-node'', where the related links are redefined as links with other super-nodes. Interdependency relations are then defined between super-nodes in different networks. This forms a new system of interdependent super-node networks. We assume that a regular node in the original networks survives if and only if it belongs to a node group that belongs to the mutual giant component of the interdependent super-node networks. For group percolation in fully interdependent networks, we can obtain the mutual giant component size of the super-node networks, as well as the size of the surviving regular nodes in the original networks (called the mutual ``giant grouped component''). To prove the non-existence of second-order percolation transitions under random initial attacks on either super-nodes or regular nodes, we present a new approach based on the Jacobian matrix. Finally, to illustrate an application of the proposed group percolation approach, we apply this framework on the percolation of interdependent networks with inter-similarity under random regular node attacks.


\begin{figure*}
\includegraphics[width=0.8\textwidth, angle = 0]{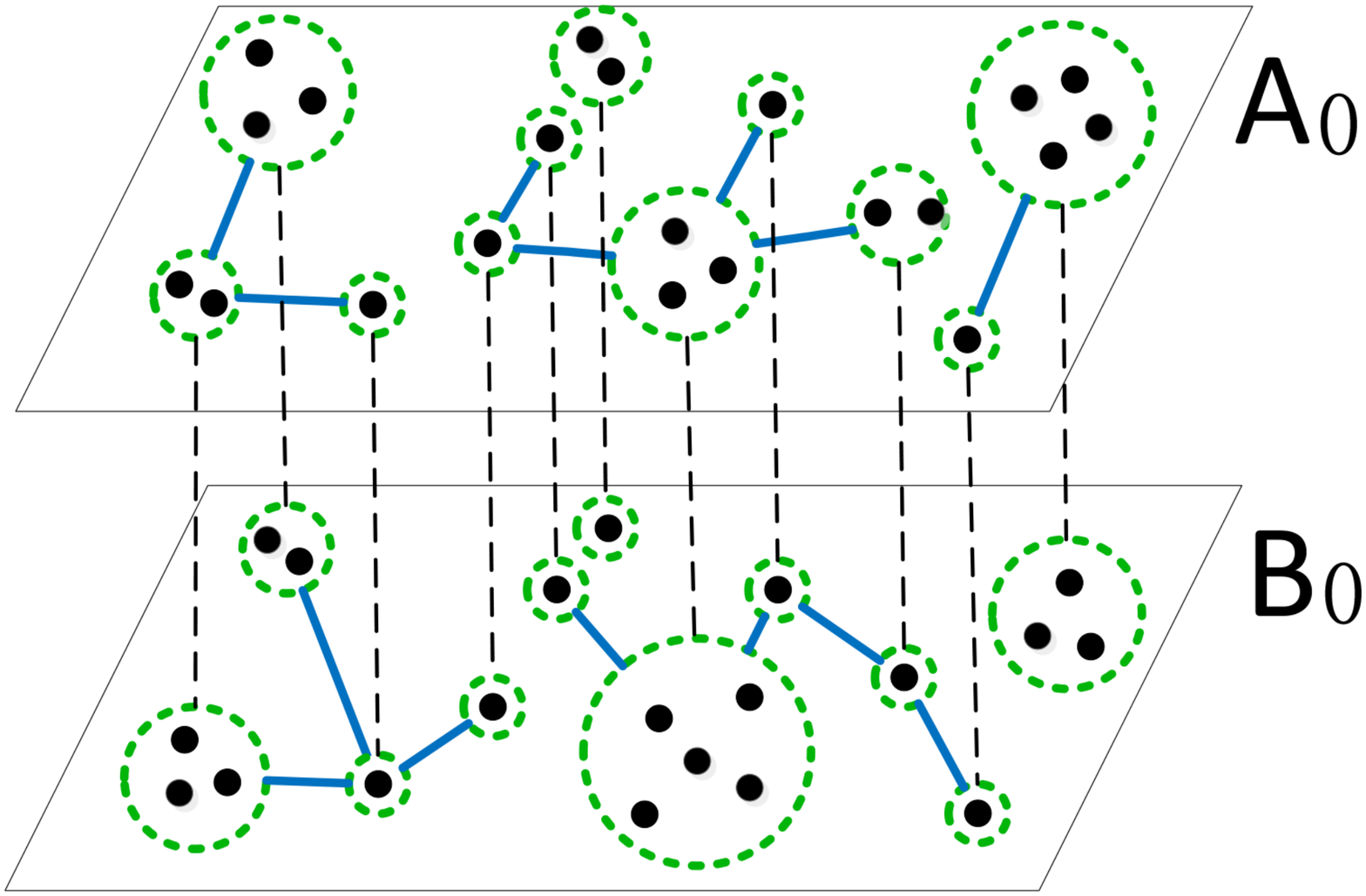}
\caption{\label{Fig1}(Color online) Demonstration of the group
percolation model. Here networks $A_0$
 and $B_0$ are the resultant networks after contracting randomly
 the nodes in networks $A$ and $B$
into super-nodes, represented as dashed circles. We assume that
networks $A_0$ and $B_0$ are
fully interdependent with the same number of super-nodes. }
\end{figure*}

\section{GENERAL MODEL}
\subsection{Probabilistic description of interdependent networks of super-nodes}
In this subsection, we introduce the basic definitions for the proposed group percolation model, where networks of super-nodes are constructed. We also develop the theoretical approach to obtain the percolation properties of the contracted interdependent networks of super-nodes.

\begin{figure*}
\includegraphics[width=0.80\textwidth,angle=0]{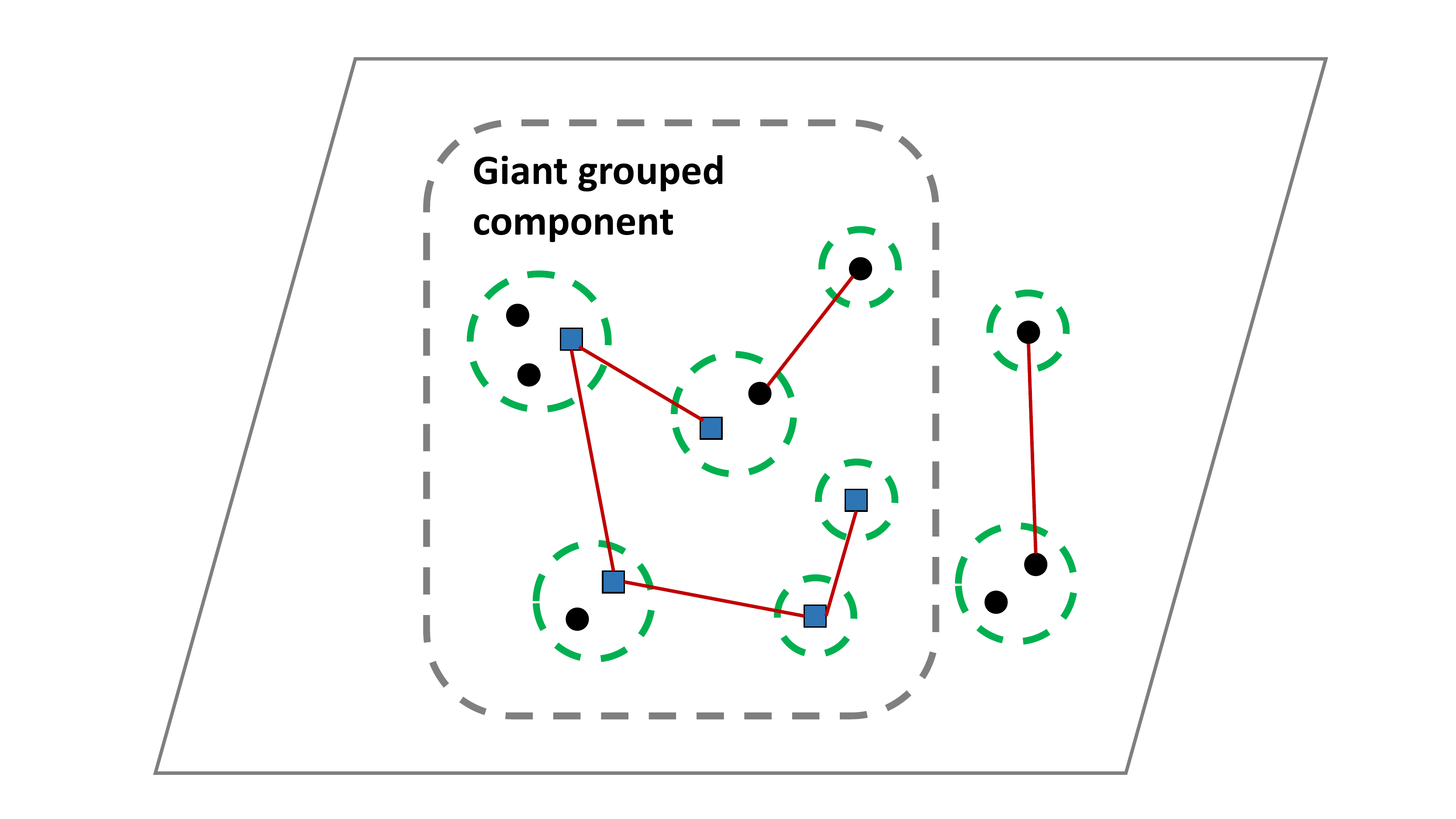}
\caption{\label{Fig1b}(Color online) A schematic of a larger giant grouped component than the giant component on a single-layer network. The red lines represent links in the original regular node network. Blue nodes (blue rectangles) constitute the giant component of the regular node network. However, all regular nodes inside the grey dashed line belong to the giant grouped component, since the corresponding super-nodes belong to the giant component of the contracted network of super-nodes. This exemplifies the situation with a larger giant grouped component than the giant component due to node group combination.}
\end{figure*}

For simplicity and without loss of generality, we consider two fully interdependent random networks $A$ and $B$ of sizes $N_A$ and $N_B$ with degree distributions $P_{A}(k)$ and $P_{B}(k)$, respectively.
We randomly divide all nodes in network $A$ into $N$ non-overlapping (without shared nodes) node groups. We do the same in network $B$, and then networks $A$ and $B$ have the same number of node groups $N$.
Next we contract each node group into a ``super-nodes'', and all the related links will be merged onto this super-node. In this way, two networks of super-nodes, $A_0$ and $B_0$ (of the same size $N$), will be constructed (see Fig.~\ref{Fig1}). Therefore, two super-nodes $a_1$ and $a_2$ are connected by a link if and only if there is at least one link between regular nodes from the two corresponding node groups. The size of each super-node is the number of regular nodes within the corresponding node group.
We assume that every super-node from network $A_0$ depends on one and only one super-node in network $B_0$, and vice versa, with the no-feedback condition. (In fact, it can be generalized to partial interdependency relations.) This means that if one super-node in $A_0$ fails, the corresponding super-node in $B_0$ will also fail, and vise versa. We focus on the mutual giant component (the largest mutually-connected component) of the interdependent super-node networks. We assume that all regular nodes within super-nodes belonging to the mutual giant component will survive together in the original regular node interdependent networks. This set of functional nodes are defined as the mutual ``giant grouped component'' of the original system. Due to the above assumptions, some surviving regular nodes may belong to different mutually-connected components in networks $A$ and $B$ originally. Therefore, the size of the mutual giant grouped component can be greater than the mutual giant component of the original networks. Fig.~\ref{Fig1b} presents an example of a single network with a larger giant grouped component than the giant component. Compared with traditional percolation models, this model is called ``group percolation'', with more robust systems via sharing connections among grouped nodes.

In this model, the degree of a super-node is the number of links that connect with other super-nodes in each network.
Randomly choosing a pair of interdependent super-nodes $a$ from $A_0$
and $b$ from $B_0$, we define $Q(m_1,m_2)$ as the probability
that super-node $a$ is of size $m_1$ and
super-node $b$ is of size $m_2$. Then we can write the generating function of $Q(m_1,m_2)$ as
\begin{equation}
D_0(x,y)=\sum_{m_1=1}^{\infty} \sum_{m_2=1}^{\infty}Q(m_1,m_2)x^{m_1}y^{m_2}.  \label{D_x_y}
\end{equation}
We further define $P(k_1,k_2)$ as the probability that super-nodes $a$ in $A_0$ and $b$ in $B_0$ have degree values $k_1$ and $k_2$ respectively. Then its generating function is
\begin{equation}
G(x,y)=\sum_{k_1}\sum_{k_2}P(k_1,k_2)x^{k_1}y^{k_2}. \label{G_x_y}
\end{equation}
Under the condition that for each super-node $a$ there is no link in the regular node network that connects the nodes belonging to $a$, $G(x,y)$ could also be written as
\begin{equation}
G(x,y)=D_0\left( G_{A}(x), G_{B}(y)\right), \label{G_x_y_1}
\end{equation}
where $G_{A}(x)=\sum_{k} P_{A}(k)x^{k}$ and $G_{B}(y)=\sum_{k} P_{B}(k) y^{k}$
are the generating functions of the degree distributions of network $A$ and network $B$, respectively. Notice that the non-existence of inner links within each super-node is usually true if the super-node size is a zero fraction of the original network, and node groups are randomly defined. In fact, when the size of each node group is a zero fraction of $N_A$ or $N_B$ as $N_A$ or $N_B$ approaches to infinity, the probability of having inner links between nodes within the same node group becomes zero in the thermodynamical limit, which is ignorable. The detailed deduction of Eq.~(\ref{G_x_y_1}) is the following:
\begin{eqnarray}
\nonumber D_0\left( G_{A}(x), G_{B}(y)\right) &=& \sum_{m_1}\sum_{m_2} Q\left(m_1,m_2\right)\cdot G_A(x)^{m_1}G_B(y)^{m_2} \\
\nonumber &=& \sum_{m_1}\sum_{m_2} Q\left(m_1,m_2\right)\cdot \left[\sum_{f_1} P_{A}(f_1)x^{f_1}\right]^{m_1}\cdot\left[\sum_{f_2} P_{B}(f_2)y^{f_2}\right]^{m_2} \\
\nonumber &=& \sum_{m_1}\sum_{m_2} Q\left(m_1,m_2\right)\cdot \sum_{k_1} q_0(k_1|m_1)x^{k_1} \cdot \sum_{k_2} q_0(k_2|m_2)y^{k_2} \\
\nonumber &=& \sum_{k_1}\sum_{k_2}\left[\sum_{m_1}\sum_{m_2} Q\left(m_1,m_2\right)q_0(k_1|m_1)q_0(k_2|m_2)\right]\cdot x^{k_1}y^{k_2} \\
\nonumber &=& \sum_{k_1}\sum_{k_2}P(k_1,k_2)x^{k_1}y^{k_2} \\
 &=& G(x,y), \label{eq3detail}
\end{eqnarray}
where $q_{0}(k|m)$ is the conditional probability that a randomly chosen
super-node has degree of $k$ given that it is of size $m$. Notice that
\begin{eqnarray}
\nonumber \left[\sum_{f_1} P_{A}(f_1)x^{f_1}\right]^{m_1} &=& \left(P_{A}(0)+P_{A}(1)\cdot x+\cdots +P_{A}(f)\cdot x^{f_1} +\cdots\right)^{m_1}\\
\nonumber &=&q_0(0|m_1)+q_0(1|m_1)\cdot x+\cdots+q_0(f_1|m_1)\cdot x^{f_1}+\cdots\\
&=&\sum_{k_1=0}^{\infty} q_{0}(k_1|m_1)\cdot x^{k_1}
\end{eqnarray}
since the coefficient of each term after the expansion is just the sum over the probabilities of all possible combinations of $m_1$ degree values that sum up to $k_1$.

We start with two interdependent networks $A_0$
and $B_0$, in which no super-nodes are removed initially. To
get the mutual connected giant cluster in networks $A_0$ and
$B_0$, we adopt the probabilistic approach and define $t_1$
as the probability that a randomly chosen link in network $A_0$
leads to the mutual giant component at one of its end. A similar quantity
$t_2$ is defined for network $B_0$. Therefore, assume now we
randomly choose a link $l$ in network $A_0$ and reach super-node
$a$ (with degree $k_1$) at one of its ends. For super-node $a$
to be part of the mutual giant component, it itself must connect with the
mutual giant component of network $A_0$ and its interdependent
counterpart $b$ (with degree $k_2$) must also connect with
the mutual giant component of network $B_0$. Computing this probability,
we can write out the self-consistent equation for $t_1$ as
\begin{equation}
t_1=\sum_{k_1}\frac{P_{A_0}(k_1)k_1}{\left\langle k_1 \right\rangle}
\left[1-(1-t_1)^{k_1-1}\right]\cdot \sum_{k_2}P_{B_0}(k_2)\left[1-(1-t_2)^{k_2}\right], \label{t_1_1}
\end{equation}
with $\frac{P_{A_0}(k_1)k_1}{\left\langle k_1 \right\rangle}$ being the
probability that a random selected link connecting with a super-node $a$ which has a degree of $k_1$, $1-(1-t_1)^{k_1-1}$
the probability that at least one of the other $k_1-1$ links of super-node
$a$ (other than the one first chosen) leads to the mutual giant component in $A_0$, $P_{B_0}(k_2)$ the probability that super-node $b$
has a degree of $k_2$, and $1-(1-t_2)^{k_2}$ the probability that at
least one of the $k_2$ links of super-node $b$ leads to the mutual giant component in $B_0$. Considering
the inter degree-degree correlation probability $P(k_1,k_2)$ between the super-nodes
in network $A_0$ and super-nodes in network $B_0$, we recast Eq.~(\ref{t_1_1})
into the general form
\begin{equation}
t_1=\sum_{k_1}\sum_{k_2}\frac{P(k_1,k_2)k_1}{\left\langle k_1 \right\rangle}\left[1-(1-t_1)^{k_1-1}\right]\cdot \left[1-(1-t_2)^{k_2}\right].\label{t_1_2}
\end{equation}
Analogously, we get the self-consistent equation for $t_2$
\begin{equation}
t_2=\sum_{k_1}\sum_{k_2}\frac{P(k_1,k_2)k_2}{\left\langle k_2 \right\rangle}\left[1-(1-t_2)^{k_2-1}\right]\cdot \left[1-(1-t_1)^{k_1}\right].\label{t_2}
\end{equation}
Using the generating function $G(x,y)$ defined in Eq.~(\ref{G_x_y}), we
transform the expressions of $t_1$ and $t_2$ into
\begin{eqnarray}
\begin{cases}
t_1=1-\frac{G_x(1-t_1,1)+G_x(1,1-t_2)-G_x(1-t_1,1-t_2)}{G_x(1,1)};\\
t_2=1-\frac{G_y(1,1-t_2)+G_y(1-t_1,1)-G_y(1-t_1,1-t_2)}{G_y(1,1)},    \label{t_1_t_2}
\end{cases}
\end{eqnarray}
where $G_x(x,y)=\frac{\partial{G(x,y)}}{\partial{x}}$ is the partial derivative of $G(x,y)$ with respect to $x$
and likewise $G_y(x,y)=\frac{\partial{G(x,y)}}{\partial{y}}$ is the partial derivative of $G(x,y)$ with
respect to $y$. Note that $t_1$ and $t_2$  could be computed using iteration
with some proper initial values of $t_1$ and $t_2$.

Accordingly, the probability $P_{\infty}^{A_0}$ that a randomly chosen
super-node in network $A_0$ belongs to the mutual giant component
is calculated as
\begin{equation}
P_{\infty}^{A_0}=\sum_{k_1}^{\infty}\sum_{k_2}^{\infty}P(k_1,k_2)\left[1-(1-t_1)^{k_1}\right] \cdot \left[1-(1-t_2)^{k_2}\right],  \label{P_infty_A0}
\end{equation}
which could be written in terms of generating function as
\begin{equation}
P_{\infty}^{A_0}=1-G(1-t_1,1)-G(1,1-t_2)+G(1-t_1,1-t_2).  \label{P_infty_A}
\end{equation}
Since networks $A_0$ and $B_0$ are fully interdependent on a one-to-one
basis and of the same size,  naturally we have
$P_{\infty}^{A_0}=P_{\infty}^{B_0}$ as determined by Eq.~(\ref{P_infty_A}).

\subsection{Mapping from super-node network back to regular node network}

In the previous subsection, we obtained the mutual giant component of the interdependent networks of super-nodes $A_0$ and $B_0$. Here, we aim to find the corresponding size of the mutual giant grouped component in $A$ and $B$ (number of surviving regular nodes). To this end, we first explore such a relation for a single layer network, and then extend that to the two-layer case.

\subsubsection{Mapping from single super-node network back to single regular node network}

We first start with a single network $A_0$ of $N$ super-nodes
constructed from a single network $A$ of $N_A$ regular nodes ($N_A \geq N$). Note that every super-node is made
up of at least one regular node from
network $A$ and assume that we have obtained the mutual giant
component of network $A_0$ as $P_{\infty}^{A_0}$. Now
we want to get the corresponding giant grouped component $P_{G}^A$
of network $A$ as well, i.e., the ratio
of the number of regular nodes contained in the super-nodes of
$P_{\infty}^{A_0}$ to the network size $N_A$.

Now denoting $\tilde{Q}(m)$ as the probability that a randomly chosen
super-node in network $A_0$ is of size $m$ (containing $m$ regular nodes),
we can write the generating function of $\tilde{Q}(m)$ as
\begin{equation}
{\tilde{D}_0}(x)=\sum_{m=1}^{\infty} \tilde{Q}(m) \cdot x^m.
\end{equation}
 Using ${\tilde{D}_0}(x)$ we further define ${T}(x)$ as
\begin{equation}
T(x)=x\cdot  \tilde{D}_0^{'}(x)=\sum_{m=1}^{\infty} m \cdot \tilde{Q}(m)\cdot  x^{m}. \label{T_x}
\end{equation}
Note that $T(1)$ is the average number of regular nodes contained in
each super-node of $A_0$.
Accordingly, recalling $G_{A}(x)=\sum_{k} P_{A}(k) x^{k}$,
we have $T(G_{A}(x))$
\begin{equation}
T(G_{A}(x))=\sum_{m=1}^{\infty} m\cdot  \tilde{Q}(m)\cdot \left[\sum_f P_{A}(f)x^f\right]^m.\\ \label{TGx}
\end{equation}
Similarly to Eq.~(\ref{eq3detail}), we know $\left[\sum_f P_{A}(f)x^f\right]^m=\sum_{k=0}^{\infty} q_{0}(k|m)\cdot x^k$. Note that
 $\sum_{k=0}^{\infty} q_{0}(k|m)=1$. Therefore, Eq.~(\ref{TGx}) can be recast into
\begin{eqnarray}
\nonumber T(G_{A}(x))&=&\sum_{m=1}^{\infty} m\cdot \tilde{Q}(m)\cdot \sum_{k=0}^{\infty} q_{0}(k|m) \cdot x^k\\ \label{TG2}
 &=&\sum_{k=0}^{\infty}\sum_{m=1}^{\infty} m \cdot \tilde{Q}(m)\cdot q_{0}(k|m) \cdot x^k.  \label{T_G_A_x}
\end{eqnarray}
Next we define $f(k,m)$ as the joint probability that a randomly chosen
super-node in network $A_0$  has degree of $k$ and is of size $m$.
Thus according to Bayes' rule,  we can write $f(k,m)$ out either as a
product of $\tilde{Q}(m)$ and $q_0(k|m)$, i.e.,
\begin{equation}
f(k,m)=\tilde{Q}(m)\cdot q_{0}(k|m),
\end{equation}
or as a product of $P_{A_0}(k)$ and $q_1(m|k)$, i.e.,
\begin{equation}
f(k,m)=P_{A_0}(k)\cdot q_{1}(m|k), \label{Bay}
\end{equation}
where $ q_{1}(m|k)$ is the conditional probability that
a randomly chosen super-node is of size $m$ given that it has
degree of $k$. Therefore using Eqs.~(\ref{T_G_A_x})-(\ref{Bay}),
 Eq.~(\ref{TGx}) can be rewritten as
\begin{eqnarray}
\nonumber T(G_{A}(x))&=&\sum_{k=0}^{\infty}\sum_{m=1}^{\infty} m\cdot \tilde{Q}(m)\cdot q_{0}(k|m)x^k\\
 &=&\sum_{k=0}^{\infty}\sum_{m=1}^{\infty} m\cdot q_{1}(m|k)\cdot P_{A_0}(k) x^k.
\end{eqnarray}
By further defining $R_{A_0}(k)$, the average size of all the super-nodes
with degree $k$, as
\begin{equation}
R_{A_0}(k)=\sum_{m=1}^{\infty} m\cdot q_{1}(m|k),
\end{equation}
we get a compact form of $T(G_{A}(x))$ as
\begin{equation}
T(G_{A}(x))=\sum_{k=0}^{\infty} R_{A_0}(k)\cdot P_{A_0}(k)\cdot x^k. \label{T_G_A}
\end{equation}

Similar to Eq.~(\ref{P_infty_A0}), we can obtain the probability that a
randomly chosen super-node from network $A_0$ is connected
to the giant component as
\begin{equation}
P_{\infty}^{A_0}=\sum_k P_{A_0}(k)[1-(1-t_1)^k].
\end{equation}
Likewise the probability that a super-node of size $m$ leads
to the giant component is $\sum_k q_0(k|m)[1-(1-t_1)^k]$. Also note that in all the super-nodes of size $m$ in network $A_0$,
there are altogether $m\cdot N\cdot {\widetilde{Q}}(m)$ regular nodes. Therefore, there are $m\cdot N\cdot {\widetilde{Q}}(m) \cdot \sum_k q_0(k|m)[1-(1-t_1)^k]$
regular nodes contained in those super-nodes of size $m$ that are connected
to the giant component. Thus taking into account all the sizes of super-nodes in the giant component, we can get the probability $P_{G}^{A}$ that a randomly chosen regular node in network $A$ belongs to the
giant grouped component
\begin{eqnarray}
\nonumber P_{G}^{A}&=&\frac{\sum_{m=1}^{\infty} m\cdot N\cdot {\widetilde{Q}}(m)\cdot\sum_k q_0(k|m)[1-(1-t_1)^k]}{NT(1)}\\
 &=&1-\frac{T(G_{A}(1-t_1))}{T(1)}. \label{tilder_P_single}
\end{eqnarray}
Note that $P_{G}^A$ is also the
normalized size of the giant grouped component in network $A$. The giant grouped component may contains more regular nodes than the giant component of network $A$, since nodes belonging to different connected components can be merged into the same super-node (once again please see Fig.~\ref{Fig1b}).

\subsubsection{Mapping from interdependent super-node networks back to interdependent regular node network}
Following the mapping strategy laid down in the previous subsection,
for fully interdependent networks of super-nodes
$A_0$ and $B_0$ we first use $D_0(x,y)$ from Eq.~(\ref{D_x_y}) to get
\begin{equation}
\begin{cases}
T_{A_0}(x,y)=x\cdot\frac{\partial D_0(x,y)}{\partial x},\\  \label{T_A_x}
T_{B_0}(x,y)=y\cdot\frac{\partial D_0(x,y)}{\partial y}.
\end{cases}
\end{equation}
Here $T_{A_0}(x,y)$ and $T_{B_0}(x,y)$ serve the same purpose of
$T(x)$ defined in Eq.~(\ref{T_x}).
Performing the derivation demonstrated in Eqs.~(\ref{TGx})-(\ref{T_G_A}),
we have $T_{A_0}(G_{A}(x), G_{B}(y))$ equal to
\begin{equation}
 T_{A_0}\left(G_{A}(x), G_{B}(y)\right)=\sum_{k_1=0}^{\infty} \sum_{k_2=0}^{\infty} R_{A_0}(k_1,k_2)\cdot P(k_1,k_2) x^{k_1} y^{k_2}, \label{T_G_A_B}
\end{equation}
where $R_{A_0}(k_1,k_2)$ is an extension of $R_{A_0}(k)$ and denotes the average number of regular nodes contained
in a super-node $a$ of degree $k_1$ in network $A_0$
with its dependency counterpart $b$ a super-node of degree $k_2$ in network $B_0$;
 $P(k_1,k_2)$ is the probability that a super-node $a$ of degree $k_1$ in
 network $A_0$ is connected by bidirectional dependency link to a super-node
 $b$ of degree $k_2$ in network $B_0$.  Analogously, using
 $P(k_1,k_2)$,  $T_{B_0}(G_{A}(x), G_{B}(y))$ takes the form of
 \begin{equation}
  T_{B_0}\left(G_{A}(x), G_{B}(y)\right)=\sum_{k_1=0}^{\infty} \sum_{k_2=0}^{\infty} R_{B_0}(k_1,k_2)\cdot P(k_1,k_2) x^{k_1} y^{k_2}, \label{T_G_A_B_0}
 \end{equation}
 with $R_{B_0}(k_1,k_2)$ the average number of regular nodes contained
 in a super-node $b$ of degree $k_2$ in network $B_0$
 with its dependency counterpart $a$ a super-node of degree $k_1$ in network $A_0$;
 In the  fully interdependent scenario, networks $A_0$ and $B_0$ are of the
 same size $N$. Following similar arguments leading up to Eq.~(\ref{tilder_P_single})
 in the previous subsection,  we get the probability $P_{G}^{A}$ that a
 randomly chosen regular node belongs to the mutual giant grouped component in network $A$ as
\begin{eqnarray}
\nonumber P_{G}^{A} &=&1-\frac{{T_{A_0}\left( G_{A}(1-t_1), G_{B}(1)\right)+T_{A_0}\left( G_{A}(1) ,G_{B}(1-t_2)\right)}}{T_{A_0}(1,1)} \\
&& +\frac{T_{A_0}\left( G_{A}(1-t_1),G_{B}(1-t_2)\right)}{T_{A_0}(1,1)}.
\end{eqnarray}
After denoting $F_A(x,y)=T_{A_0}(G_{A}(x), G_{B}(y))/T_{A_0}(1,1)$,
we can simplify the above equation to
\begin{equation}
P_{G}^{A} =1-F_A(1-t_1,1)-F_A(1,1-t_2)+F_A(1-t_1,1-t_2).\label{P_inf_A}
\end{equation}
Analogously, we can write out $P_{G}^{B}$ as
\begin{equation}
 P_{G}^{B}=1-F_B(1-t_1,1)-F_B(1,1-t_2)+F_B(1-t_1,1-t_2), \label{P_inf_B}
\end{equation}
with $F_B(x,y)=T_{B_0}(G_{A}(x), G_{B}(y))/T_{B_0}(1,1)$. Note that
$P_{G}^{A}$ ($P_{G}^{B}$) is also the
normalized size of the mutual giant grouped component in network $A$ ($B$).

\subsection{Fully interdependent networks of super-nodes under random super-node removal}
Previous subsections have laid out the formalism of calculating the size of both the giant component of super-node networks and the giant grouped component in the corresponding regular node networks with full occupation, i.e., every super-node is considered functioning before the cascading process begins. In this subsection, we consider the case where a fraction $1-r$ of super-nodes from network $A_0$ are randomly removed to initiate the iterative cascading failure process in the interdependent networks of super-nodes. When no more super-nodes fail, networks $A_0$ and $B_0$ reach their final steady state. As a result of the initial random removal of super-nodes, $t_1$ and $t_2$ defined in Eq.~(\ref{t_1_t_2}) are slightly modified to
\begin{eqnarray}
\begin{cases}
t_1=r\cdot \left[1-\frac{G_x(1-t_1,1)+G_x(1,1-t_2)-G_x(1-t_1,1-t_2)}{G_x(1,1)}\right], \\\label{new_t_1_t_2}
t_2=r\cdot \left[1-\frac{G_y(1,1-t_2)+G_y(1-t_1,1)-G_y(1-t_1,1-t_2)}{G_y(1,1)}\right].
\end{cases}
\end{eqnarray}
Likewise, $P_{G}^{A}$ defined in Eq.~(\ref{P_infty_A}) will be
correspondingly modified to
\begin{equation}
P_{G}^{A}=r \cdot \left[1-G(1-t_1,1)-G(1,1-t_2)+G(1-t_1,1-t_2)\right].  \label{p_P_infty}
\end{equation}
Therefore, as a function of $t_1$, $t_2$ and $p$, at the steady state, the probability $P_{G}^{A}$ ($P_{G}^{B}$) that a randomly chosen regular node in network $A$ ($B$) belongs to the mutual giant grouped component can also be tailored from Eqs.~(\ref{P_inf_A}) and (\ref{P_inf_B}) as
\begin{eqnarray}
\begin{cases}
 P_{G}^{A}=r\cdot \left[1-F_A(1-t_1,1)-F_A(1,1-t_2)+F_A(1-t_1,1-t_2)\right], \\ \label{P_A_B_p}
P_{G}^{B}=r\cdot \left[1-F_B(1-t_1,1)-F_B(1,1-t_2)+F_B(1-t_1,1-t_2) \right].
\end{cases}
\end{eqnarray}

\begin{figure*}
 \centering
 \subfigure{
  \label{fig_model1a}
  \includegraphics[width=0.46\textwidth]{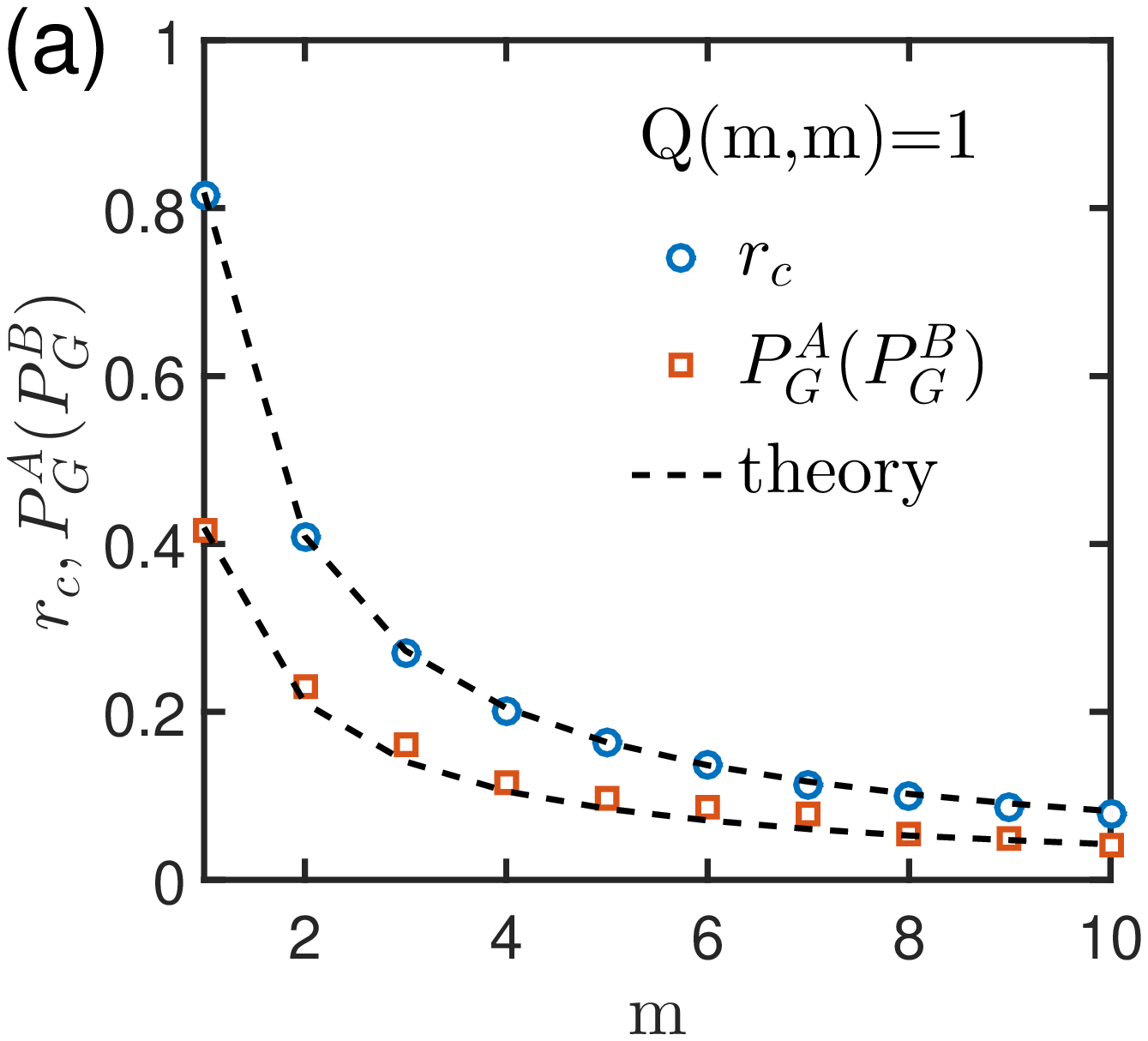}}
 \subfigure{
  \label{fig_model1b}
  \includegraphics[width=0.46\textwidth]{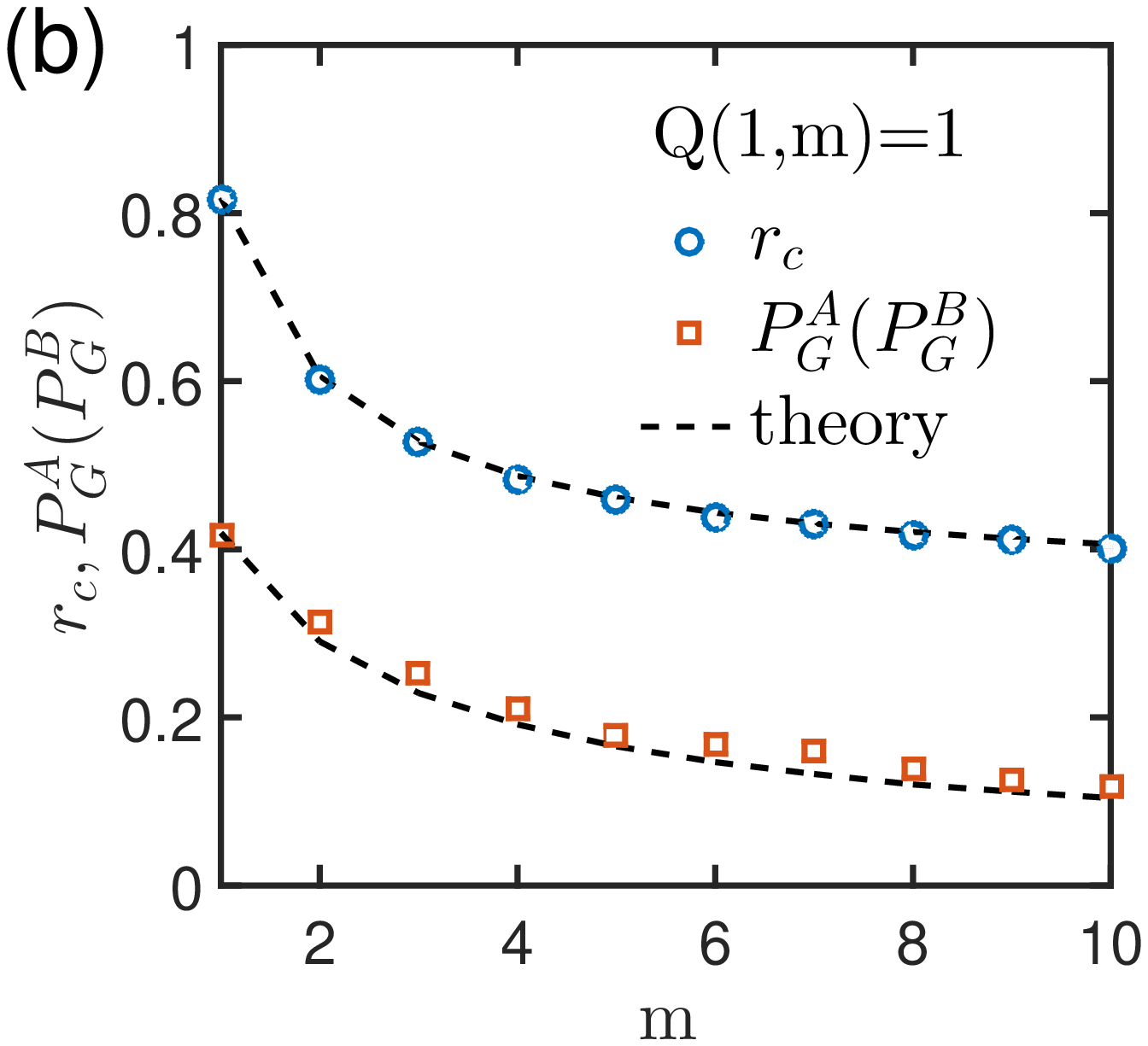}}
 \caption{\label{Fig2}(Color online) (a) Transition point $r_c$ and the mutual giant grouped component size at $r_c$ (the jump size) as functions of $m$, with a super-node size distribution $Q(m,m)=1$, $m=1,2,\cdots,10$. (b) The same with (a) but with $Q(1,m)=1$, $m=1,2,\cdots,10$. 
 Networks $A_0$ and $B_0$ are obtained from two ER networks $A$ and $B$ with $\left\langle k_{A}\right\rangle=\left\langle k_{B} \right\rangle=3$. Dashed-lines are theoretical predictions and symbols are simulation results. Network sizes vary for different $m$ values in order to keep the number of super-nodes $N=5000$.
}
\end{figure*}

The group percolation results under random super-node initial removal are shown in Fig.~\ref{Fig2}. Here we show two different super-node size distributions based on two ER networks: $Q(m,m)=1$ and $Q(1,m)=1$, where $m$ ranges from 1 to 10. We plot the transition point $r_c$ as well as the mutual giant grouped component size at $r_c$ (the jump size) as functions of $m$. We find that in both cases, simulation results exhibit first-order percolation transitions, and they agree with the analytical prediction very well.

Next we prove that the fully interdependent networks of super-nodes will undergo first-order phase transitions under random super-node removal. One trivial solution of Eq.~(\ref{new_t_1_t_2}) is $\bigl( \begin{smallmatrix} t_1 \\ t_2 \end{smallmatrix} \bigr)=\bigl( \begin{smallmatrix} 0 \\ 0 \end{smallmatrix} \bigr)$, which corresponds to a mutual giant grouped component with size $0$. We assume that when $r$ reaches a certain critical value $r_c$, a non-zero solution of Eq.~(\ref{new_t_1_t_2}), $\bigl( \begin{smallmatrix} t_1 \\ t_2 \end{smallmatrix} \bigr)=\bigl( \begin{smallmatrix} \tilde{t}_1 \\ \tilde{t}_2\end{smallmatrix} \bigr)$ appears and increases from $\bigl( \begin{smallmatrix} 0 \\ 0 \end{smallmatrix} \bigr)$ continuously, which indicates that when $r\rightarrow r_c$ from above, $\bigl( \begin{smallmatrix} \tilde{t}_1 \\ \tilde{t}_2 \end{smallmatrix} \bigr)\rightarrow\bigl( \begin{smallmatrix} 0 \\ 0 \end{smallmatrix} \bigr)$. We denote the left and the right hand sides of Eq.~(\ref{new_t_1_t_2}) as ${\rm{\mathbf{\Phi}}}_{1}({\rm{\mathbf{t}}})=\left(t_1,t_2\right)^{T}$ and ${\rm{\mathbf{\Phi}}}_{2}({\rm{\mathbf{t}}})=\left(r\cdot \left[1-\frac{G_x(1-t_1,1)+G_x(1,1-t_2)-G_x(1-t_1,1-t_2)}{G_x(1,1)}\right],r\cdot\left[1-\frac{G_y(1,1-t_2)+G_y(1-t_1,1)-G_y(1-t_1,1-t_2)}{G_y(1,1)}\right]\right)^{T}$ respectively. Therefore, we have ${\rm{\mathbf{\Phi}}}_{1}\bigl( \begin{smallmatrix} 0 \\ 0 \end{smallmatrix} \bigr)={\rm{\mathbf{\Phi}}}_{2}\bigl( \begin{smallmatrix} 0 \\ 0 \end{smallmatrix} \bigr)=\bigl( \begin{smallmatrix} 0 \\ 0 \end{smallmatrix} \bigr)$, and ${\rm{\mathbf{\Phi}}}_{1}\bigl( \begin{smallmatrix} \tilde{t}_1 \\ \tilde{t}_2 \end{smallmatrix} \bigr)={\rm{\mathbf{\Phi}}}_{2}\bigl( \begin{smallmatrix} \tilde{t}_1 \\ \tilde{t}_2 \end{smallmatrix} \bigr)$. Notice that ${\rm{\mathbf{\Phi}}}_{1}({\rm{\mathbf{t}}})$ and ${\rm{\mathbf{\Phi}}}_{2}({\rm{\mathbf{t}}})$ are both derivable at $\bigl( \begin{smallmatrix} 0 \\ 0 \end{smallmatrix} \bigr)$, thus they can be expressed as

\begin{equation}
\begin{array}{c}
{\rm{\mathbf{\Phi}}}_{1}\left(\begin{array}{c} t_1 \\ t_2 \end{array}\right) = {\rm{\mathbf{\Phi}}}_{1}\left(\begin{array}{c} 0 \\ 0 \end{array}\right) + {\rm{\mathbf{J}}}_{{\rm{\mathbf{\Phi}}}_{1}}\bigg|_{\bigl( \begin{smallmatrix} t_1 \\ t_2 \end{smallmatrix} \bigr)=\bigl( \begin{smallmatrix} 0 \\ 0 \end{smallmatrix} \bigr)}\cdot\left(\begin{array}{c} t_1 \\ t_2 \end{array}\right) + o\left(\begin{array}{c} t_1 \\ t_2 \end{array}\right), \\ {\rm{\mathbf{\Phi}}}_{2}\left(\begin{array}{c} t_1 \\ t_2 \end{array}\right) = {\rm{\mathbf{\Phi}}}_{2}\left(\begin{array}{c} 0 \\ 0 \end{array}\right) + {\rm{\mathbf{J}}}_{{\rm{\mathbf{\Phi}}}_{2}}\bigg|_{\bigl( \begin{smallmatrix} t_1 \\ t_2 \end{smallmatrix} \bigr)=\bigl( \begin{smallmatrix} 0 \\ 0 \end{smallmatrix} \bigr)}\cdot\left(\begin{array}{c} t_1 \\ t_2 \end{array}\right) + o\left(\begin{array}{c} t_1 \\ t_2 \end{array}\right),
\end{array}
\end{equation}

\noindent where ${\rm{\mathbf{J}}}_{{\rm{\mathbf{\Phi}}}_{1}}\big|_{\bigl( \begin{smallmatrix} t_1 \\ t_2 \end{smallmatrix} \bigr)=\bigl( \begin{smallmatrix} 0 \\ 0 \end{smallmatrix} \bigr)}=\bigl( \begin{smallmatrix} 1 & 0 \\ 0 & 1 \end{smallmatrix} \bigr)$ and ${\rm{\mathbf{J}}}_{{\rm{\mathbf{\Phi}}}_{2}}\big|_{\bigl( \begin{smallmatrix} t_1 \\ t_2 \end{smallmatrix} \bigr)=\bigl( \begin{smallmatrix} 0 \\ 0 \end{smallmatrix} \bigr)}=\bigl( \begin{smallmatrix} 0 & 0 \\ 0 & 0 \end{smallmatrix} \bigr)$ are the Jacobian matrices of ${\rm{\mathbf{\Phi}}}_{1}$ and ${\rm{\mathbf{\Phi}}}_{2}$ at $\bigl( \begin{smallmatrix} 0 \\ 0 \end{smallmatrix} \bigr)$ respectively, and $o\bigl( \begin{smallmatrix} t_1 \\ t_2 \end{smallmatrix} \bigr)$ denotes a multivariable higher-order infinitesimal of $\bigl( \begin{smallmatrix} t_1 \\ t_2 \end{smallmatrix} \bigr)$. When the non-zero solution $\bigl( \begin{smallmatrix} \tilde{t}_1 \\ \tilde{t}_2 \end{smallmatrix} \bigr)$ approaches $\bigl( \begin{smallmatrix} 0 \\ 0 \end{smallmatrix} \bigr)$, we have

\begin{equation}
\begin{array}{c}
{\rm{\mathbf{\Phi}}}_{1}\left(\begin{array}{c} \tilde{t}_1 \\ \tilde{t}_2 \end{array}\right) = {\rm{\mathbf{\Phi}}}_{1}\left(\begin{array}{c} 0 \\ 0 \end{array}\right) + \left(\begin{matrix} 1 & 0 \\ 0 & 1 \end{matrix}\right)\cdot\left(\begin{array}{c} \tilde{t}_1 \\ \tilde{t}_2 \end{array}\right) + o\left(\begin{array}{c} \tilde{t}_1 \\ \tilde{t}_2 \end{array}\right)=\left(\begin{array}{c} \tilde{t}_1 \\ \tilde{t}_2 \end{array}\right) + o\left(\begin{array}{c} \tilde{t}_1 \\ \tilde{t}_2 \end{array}\right), \\ {\rm{\mathbf{\Phi}}}_{2}\left(\begin{array}{c} \tilde{t}_1 \\ \tilde{t}_2 \end{array}\right) = {\rm{\mathbf{\Phi}}}_{2}\left(\begin{array}{c} 0 \\ 0 \end{array}\right) + \left(\begin{matrix} 0 & 0 \\ 0 & 0 \end{matrix}\right)\cdot\left(\begin{array}{c} \tilde{t}_1 \\ \tilde{t}_2 \end{array}\right) + o\left(\begin{array}{c} \tilde{t}_1 \\ \tilde{t}_2 \end{array}\right)=\left(\begin{array}{c} 0 \\ 0 \end{array}\right) + o\left(\begin{array}{c} \tilde{t}_1 \\ \tilde{t}_2 \end{array}\right).
\end{array} \label{Phi_1_Phi_2}
\end{equation}

\noindent According to ${\rm{\mathbf{\Phi}}}_{1}\bigl( \begin{smallmatrix} \tilde{t}_1 \\ \tilde{t}_2 \end{smallmatrix} \bigr)={\rm{\mathbf{\Phi}}}_{2}\bigl( \begin{smallmatrix} \tilde{t}_1 \\ \tilde{t}_2 \end{smallmatrix} \bigr)$, this leads to a contradiction: $\bigl( \begin{smallmatrix} \tilde{t}_1 \\ \tilde{t}_2 \end{smallmatrix} \bigr)=o\bigl( \begin{smallmatrix} \tilde{t}_1 \\ \tilde{t}_2 \end{smallmatrix} \bigr)$ (This means $\left| (\tilde{t}_1, \tilde{t}_2)^T \right| = \left| o(\tilde{t}_1, \tilde{t}_2)^T \right| = o \left( \left| (\tilde{t}_1, \tilde{t}_2)^T \right| \right)$, which cannot happen). Therefore, the previous assumption that the non-zero solution $\bigl( \begin{smallmatrix} t_1 \\ t_2 \end{smallmatrix} \bigr)$ occurs continuously is not true. Since the mutual giant grouped component sizes $P_{G}^{A}$ and $P_{G}^{B}$ are continuous functions of $t_1$ and $t_2$, the occurrence of the mutual giant grouped component in the regular node networks must be also discontinuous. This proves the existence of first-order percolation transitions.

The approach we used here to prove the non-existence of second-order transitions is a two-dimensional extension of the one-dimensional case discussed in \cite{hu2013percolation}, where the two functions intersect at $1$, but they have different derivatives at $1$. G. J. Baxter \textit{et al.} have also used Jacobian matrices to find the percolation-like transition points in interdependent networks \cite{BaxterPRL2012}.

\begin{figure*}
\includegraphics[width=0.8\textwidth, angle = 0]{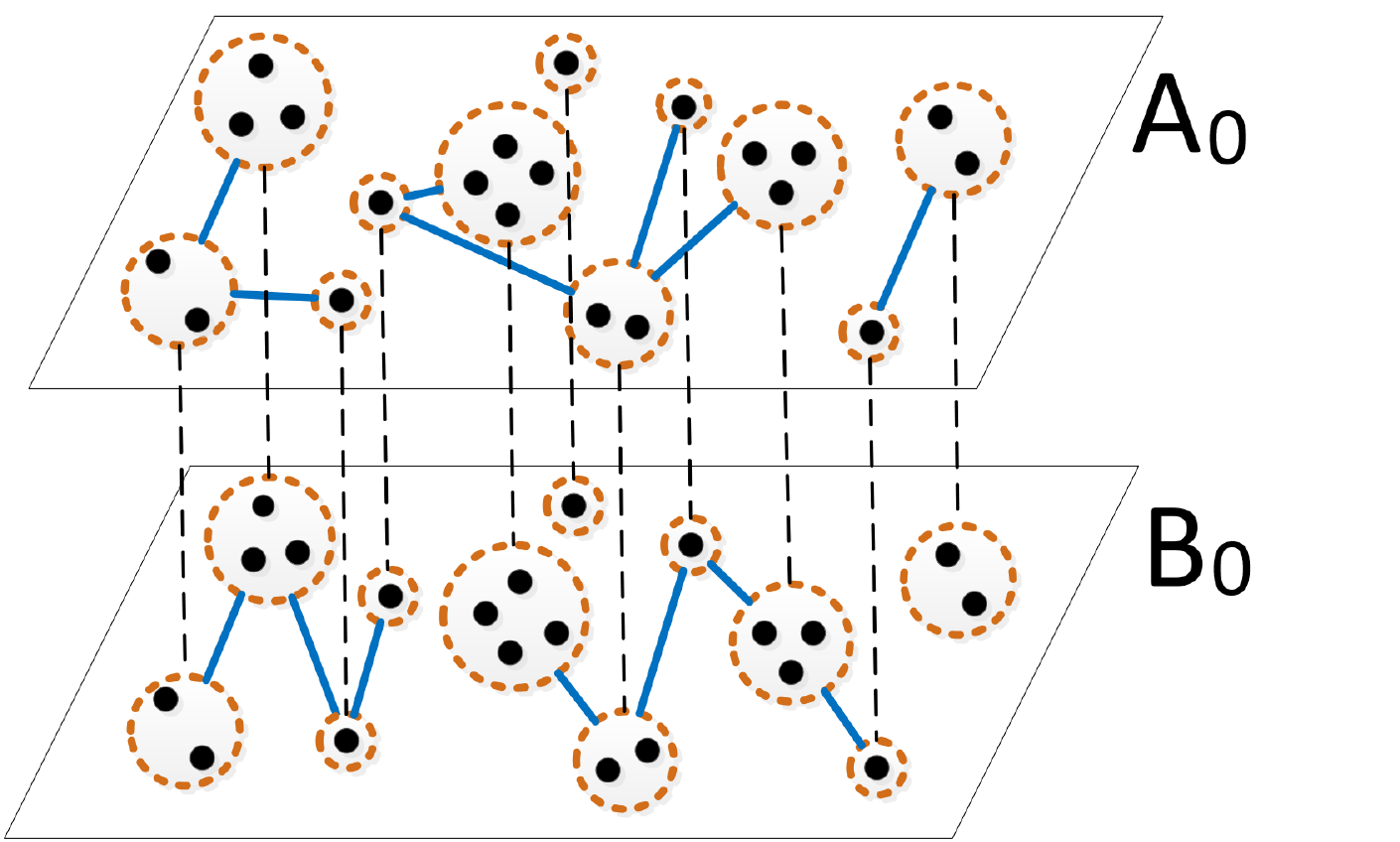}
\caption{\label{Fig3}(Color online) Demonstration of the special case of group
percolation model where networks $A_0$ and $B_0$ have
the same number of super-nodes and every super-node in network $A_0$ is
connected to a same sized super-node
in network $B_0$ via a bidirectional dependency link. }
\end{figure*}

\subsection{Fully interdependent networks of super-nodes under random regular node removal}

In the previous subsection, we have studied the group percolation after random super-node attacks. In fact, our model also allows to explore the group percolation after random regular node removal, where node groups are defined after the initial regular node attacks. In some real-world applications the initial attacks are upon regular nodes, while in some other examples the initial attacks are on groups of nodes or super-nodes. In this sense, this model can be widely applied in different types of coupled systems.

Here we consider two fully interdependent networks $A$ and $B$, with the same number of nodes $N_A=N_B=\tilde{N}$. Assume that at the beginning a fraction $1-p$ of regular nodes are randomly attacked in network $A$. According to the dependency relations, the corresponding $(1-p)\tilde{N}$ nodes in networks $B$ will also fail. We denote the attacked networks $A^\prime$ and $B^\prime$. After the initial attack, we again define the node groups in both networks and merge them into super-nodes.

Here we focus on the following special case of group percolation: each pair of interdependent super-nodes $a$ in the contracted network $A_0$ and $b$ in network $B_0$ have the same size (as shown in Fig.~\ref{Fig3}). This special case is considered here since in this way, the contracted networks $A_0$ and $B_0$ still have well-defined one-to-one full interdependency links. Moreover, in the next section, we will present the application of the proposed group percolation approach on interdependent networks with inter-similarity, where super-nodes with interdependencies always have the same size (see \cite{hu2013percolation}). Therefore, here we focus on this special case.

To analyse this case, we need to provide new degree distributions of networks $A^\prime$ and $B^\prime$. For network $A^\prime$, after the initial attack, the degree distribution is
\begin{equation}
P_{A^\prime}(k)=\sum\limits^{\infty}_{k_0=k}P_{A}(k_0)\cdot {k_0 \choose k}\cdot p^{k}\cdot (1-p)^{k_0-k}.
\end{equation}
Therefore, the corresponding generating function becomes
\begin{equation}
G_{A^\prime}(x)=\sum\limits_{k=0}^{\infty}\sum\limits^{\infty}_{k_0=k}P_{A}(k_0) {k_0 \choose k}p^{k}(1-p)^{k_0-k}x^{k},
\end{equation}
which finally becomes (following previous works on epidemic models, see Eq. (13) in \cite{NewmanPRE2002})
\begin{equation}
G_{A^\prime}(x)=\sum\limits_{k=0}^{\infty}P_{A}(k_0)\cdot(1-p+xp)^{k}=G_{A}(1+(x-1)p).
\end{equation}
Similarly, for network $B^\prime$ the generating function of the degree distribution is $G_{B^\prime}(x)=G_{B}(1+(x-1)p)$.

Since the two contracted networks $A_0$ and $B_0$ have the same super-node sizes, we denote $Q(m)$ the probability that a randomly chosen super-node in network $A_0$ (or $B_0$) is of size $m$. Then we can simplify $D_0(x,y)$ defined in Eq.~(\ref{D_x_y}) to
\begin{equation}
D_0(x)=\sum_{m=1}^{\infty}Q(m)\cdot x^m. \label{D_0_x}
\end{equation}
Similarly, $G(x,y)$ defined in Eq.~(\ref{G_x_y_1}) can be simplified to
\begin{equation}
G(x,y)=D_0\left(G_{A^\prime}(x)\cdot G_{B^\prime}(y)\right).  \label{Gxy_new}
\end{equation}

From the derivation performed from the above subsections, we can conclude that the probabilities $t_1$ and $t_2$ take the same form as defined in Eq.~(\ref{t_1_t_2}). Similarly to Eq.~(\ref{P_inf_A}), the relative mutual giant grouped component size in the regular node network $A^\prime$ (after initial attacks) is
\begin{equation}
P_{G}^{A^\prime}=1-F_{A^\prime}(1-t_1,1)-F_{A^\prime}(1,1-t_2)+F_{A^\prime}(1-t_1,1-t_2)  \label{Pinf_Aprime}
\end{equation}
where $F_{A^\prime}(x,y)=T_{A_0}(G_{A^\prime}(x),G_{B^\prime}(y))/T_{A_0}(1,1)$.
 The mutual giant grouped component size $P_{G}^{B^\prime}$ in network $B^\prime$ can be obtained similarly. Considering that $A_0$ and $B_0$ have the same super-node sizes in this case. Therefore, we have $P_{G}^{A^\prime}=P_{G}^{B^\prime}$. Thus, the relative mutual giant grouped component size in the original networks $A$ ($B$) is

\begin{equation}
P_{G}^{A}=P_{G}^{B}=p\cdot\left(1-F_{A^\prime}(1-t_1,1)-F_{A^\prime}(1,1-t_2)+F_{A^\prime}(1-t_1,1-t_2)\right)  \label{Pinf_AB}
\end{equation}

Finally, using exactly the same method with the case under initial super-node attacks, for regular node initial attacks we can also prove the non-existence of second-order percolation transitions. The only difference is that  the right hand side of Eq.~(\ref{new_t_1_t_2}) now has the form of the right hand side of Eq.~(\ref{t_1_t_2}), with $G(x,y)$ defined in Eq.~(\ref{Gxy_new}). It is easy to confirm that the proof in Eq.~(\ref{Phi_1_Phi_2}) in the previous subsection is still valid here.

\section{Interdependent networks with inter-similarity}

\begin{figure*}
\includegraphics[width=1.0\textwidth, angle = 0]{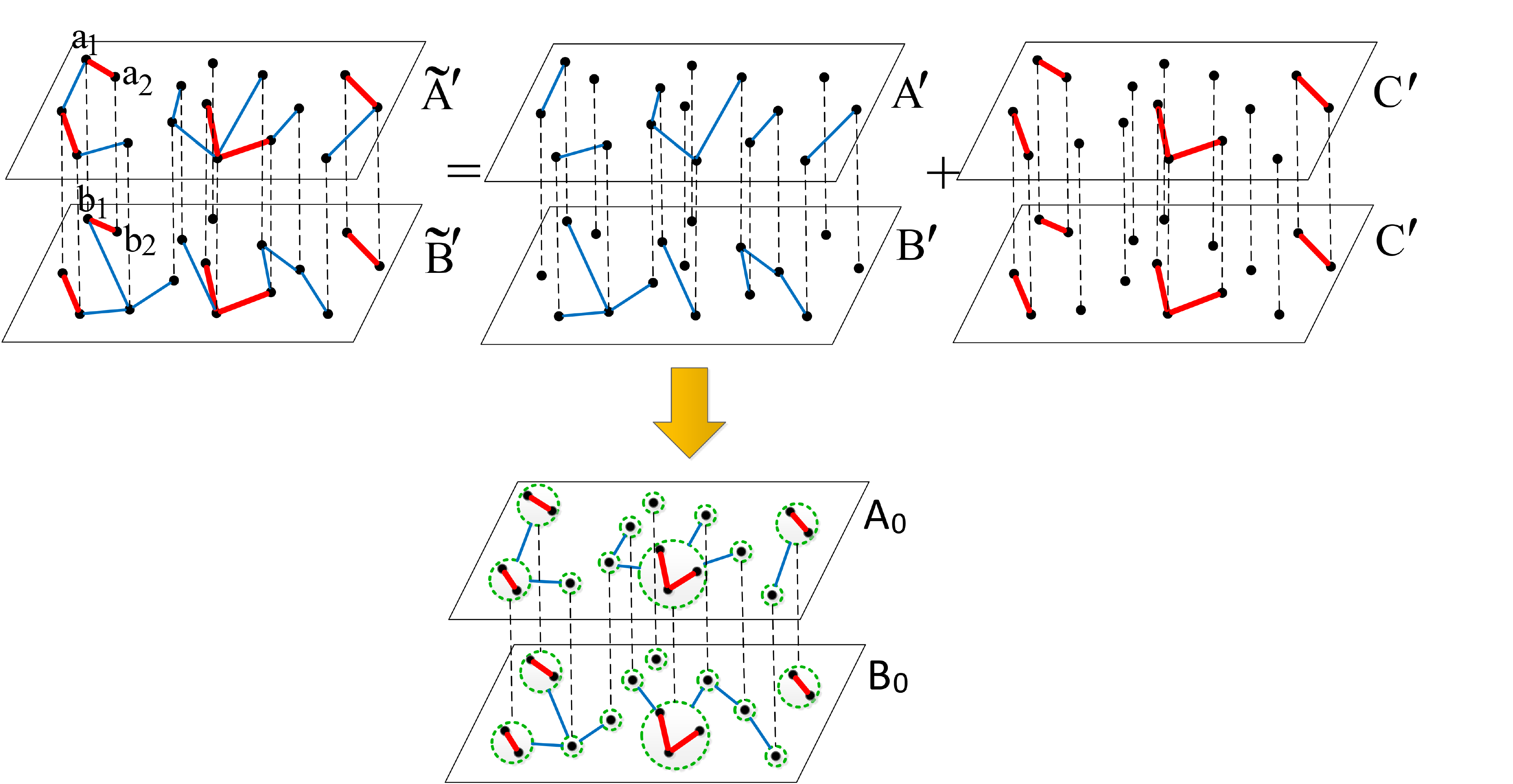}
\caption{\label{Fig4}(Color online) Decomposition of networks
$\tilde{A}^\prime$ and $\tilde{B}^\prime$ according to the common links. Bidirectional dependency links are represented by dashed lines. Solid red lines are the common connectivity links while solid blue lines are uncommon connectivity links within each network. Network $C^\prime$ contains all the common connectivity links and nodes of networks $\tilde{A^\prime}$ and $\tilde{B^\prime}$. Networks $A^\prime$ and $B^\prime$ contain all the uncommon connectivity links and nodes of networks $\tilde{A^\prime}$ and $\tilde{B^\prime}$, respectively. The connectivity link between nodes $a_1$ and $a_2$ in network $\tilde{A^\prime}$ and the connectivity link between nodes $b_1$ and $b_2$ in network $\tilde{B^\prime}$ are common connectivity links because nodes $a_1$ and $a_2$ depend on nodes $b_1$ and $b_2$, respectively. In networks $A^\prime$ and $B^\prime$, all nodes that belong to the same connected component in $C^\prime$ are grouped, and then merged into a super-node (dashed-circles) in networks $A_0$ and $B_0$. Since node groups are defined in the same way in both networks, the interdependency links are still well defined between super-nodes in $A_0$ and $B_0$.}
\end{figure*}

In the previous section, the general model of grouped percolation in interdependent networks has been introduced, and its percolation transition has been obtained analytically. In order to support the validity of these analyses, here we show the application of the group percolation approach on interdependent networks with inter-similarity. Inter-similarity means that a pair of coupled nodes have neighbors in both networks that are also coupled. This can be measured by the fraction of common links (also known as overlapping links) in interdependent or multiplex networks. The common links are defined as: given two interdependent networks $A_0$ and $B_0$, and two nodes $a_k$ and $a_l$ which are linked in $A_0$, if their interdependent counterparts $b_k$ (corresponds to $a_k$ ) and $b_l$ (corresponds to $a_l$ ) in $B_0$ are also linked, this pair of links (red links in Fig.\ref{Fig4}) is called a common link.

In fact, in one previous work (\cite{hu2013percolation}), the approach of defining super-nodes has been proposed for the first time to study the percolation of interdependent networks with inter-similarity. In that model, each super-node has the same size with its dependency counterpart. Here, we provide a much more generalized framework of group percolation, where node groups are defined independently in two networks. Therefore, here we show that the previous model in \cite{hu2013percolation} is a special case within the general framework of group percolation.

To this end, we investigate the percolation properties of two fully interdependent networks $\tilde{A}$ and $\tilde{B}$ of the same size $\tilde{N}$ with inter-similarity, where nodes in each network are randomly connected with the same degree distribution $P(k)$. Every node in network $\tilde{A}$ depends on a random node in network $\tilde{B}$, and vice versa. We also assume that if a node $i$ in network $\tilde{A}$ depends on a node $j$ in network $\tilde{B}$ and node $j$ in network $\tilde{B}$ depends on node $l$ in network $\tilde{A}$, then $l = i$, which rules out the feedback condition. This full interdependency means that every node $i$ in network $\tilde{A}$ has a dependent node $j$ in network $\tilde{B}$, and if node $i$ fails node $j$ will also fail, and vice versa.

In order to study the effects of inter-similarity (common links) on the system robustness, we assume that $\tilde{A}$ and $\tilde{B}$ can be expressed as
\begin{align}
\tilde{A}=A+C;
\quad \tilde{B}=B+C,
\end{align}
where $A$ and $B$ denote the networks with the same sets of nodes but only the non-common links; $C$ is the network including the same set of nodes and all common links between $\tilde{A}$ and $\tilde{B}$.

In such a system, initially a fraction $1-p$ of nodes in $\tilde{A}$ are randomly attacked; the corresponding nodes in $\tilde{B}$ are also removed. For the attacked networks $\tilde{A}^\prime$ and $\tilde{B}^\prime$, they can be described as
\begin{align}
\tilde{A}^\prime=A^\prime+C^\prime;
\quad \tilde{B}^\prime=B^\prime+C^\prime.
\end{align}
where $A^\prime$, $B^\prime$ and $C^\prime$ are all the remaining networks after the initial attack (see Fig.~\ref{Fig4} for illustration).

Note that nodes belonging to the same connected component in $C^\prime$ must survive in the mutual giant component of $\tilde{A}^\prime$ and $\tilde{B}^\prime$ or not, as a group (see \cite{hu2013percolation}). This is because these nodes are already mutually connected in $\tilde{A}^\prime$ and $\tilde{B}^\prime$.
This percolation can be described in the framework of group percolation. To this end, we identify and merge the corresponding components of $C^\prime$ within $A^\prime$ and $B^\prime$ into super-nodes respectively. As depicted in Fig.~\ref{Fig4}, we obtain a pair of interdependent networks $A_0$ and $B_0$ composed of super-nodes contracted from clusters spanned by common links. Moreover, every super-node $a$ in network $A_0$ is interdependent on a same sized super-node $b$ of network $B_0$, on a one-to-one correspondence basis. This is exactly the case studied in the previous subsection, and the mutual giant grouped component in this group percolation will be the same as the mutual giant component of the original attacked networks $\tilde{A}^\prime$ and $\tilde{B}^\prime$.

In order to get the giant grouped component size $P_{G}^{A^\prime}$ defined in
Eq.~(\ref{Pinf_Aprime}) for $A^\prime$ at the steady state, we need to compute $Q(m)$ first.

Next we define $k_C$ as the average degree of network $C$,
and $P_{C}(k)$
as its degree distribution. Then we can write out the generation
function of $P_{C}(k)$ as
$G_{C}(x)=\sum_{k}P_{C}(k)x^{k}$. After the initial attack, the similar generating function for $C^\prime$ is $G_{0}(x)=\sum_{k}P_{C^\prime}(k)x^{k}=G_{C}(1+(x-1)p)$. Analogously, for $C^\prime$ the generating
function of the underlying
branching process is $G_{1}(x)={G^\prime_{0}(x)}/{G^\prime_{0}(1)}$.
Moreover, the generating function
of the cluster size distributions \cite{NewmanPRE2001} by randomly traversing a link
in network $C^\prime$ is
\begin{equation}
H_1(x)=x G_1[H_1(x)],
\end{equation}
and the generating function of the cluster size distributions by
randomly traversing a super-node in network $C^\prime$ is
\begin{equation}
H_0(x)=xG_0[H_1(x)].
\end{equation}
 With $H_0(x)$ we define further $\tilde{H}_0(x)=\frac{H_0(x)}{x}$
 and get $\tilde{D}_0(x)$ as
\begin{equation}
\tilde{D}_0(x)=\int \tilde{H}_0(x) dx.
\end{equation}
Therefore $D_0(x)$ is obtained through the normalization of $\tilde{D}_0(x)$, i.e.,
$D_0(x)=\frac{\tilde{D}_0(x)}{\tilde{D}_0(1)}$. Note that $D_0(x)$
 is essentially the generating function of
the distribution $Q(m)$ as defined in Eq.~(\ref{D_0_x}).

Considering the average degree of network $C^\prime$, which is $k_{C^\prime}=pk_C$. If $k_{C^\prime} < 1$, we finally obtain the mutual giant grouped component sizes in networks $A$ and $B$, $P_{G}^{A}$ and $P_{G}^{B}$, as shown in Eq.~(\ref{Pinf_AB}). Note that they are equal to the mutual giant component sizes in the original networks $\tilde{A}$ and $\tilde{B}$: $P_{\infty}^{\tilde{A}}=P_{G}^{A}$, $P_{\infty}^{\tilde{B}}=P_{G}^{B}$.
However, if $k_{C^\prime} \geq 1$, the system
 of $A_0$ and $B_0$ would have a non-zero giant component denoted
 as $S$, which satisfies
\begin{equation}
\begin{cases}
S=1-G_{0}(u),\\
u=H_1(1).
\end{cases}
\end{equation}

In this case, we just need to slightly modify $t_1$ and $t_2$ to the following form:
\begin{eqnarray}
\begin{cases}
t_1=(1-S)\cdot \left[ 1-\frac{G_x(1-t_1,1)+G_x(1,1-t_2)-G_x(1-t_1,1-t_2)}{G_x(1,1)}\right]+ S,\\
 t_2=(1-S)\cdot\left[1-\frac{G_y(1,1-t_2)+G_y(1-t_1,1)-G_y(1-t_1,1-t_2)}{G_y(1,1)}\right]+S.
 \end{cases}
 \end{eqnarray}
The equivalent equations have been got in our previous work \cite{hu2013percolation} with $S=0$ and the work of Byungjoon {\it et al.} \cite{min2015link} by other approaches. Using $t_1$ and $t_2$ we are able to determine the size of the mutual giant
component in the original networks $\tilde{A}$ and $\tilde{B}$ as
\begin{eqnarray}
P_{\infty}^{\tilde{A}}=P_{\infty}^{\tilde{B}} = p\cdot \left( (1-S)\left[1-F_{\tilde{A}^\prime}(1-t_1,1)-F_{\tilde{A}^\prime}(1,1-t_2)+F_{\tilde{A}^\prime}(1-t_1,1-t_2)\right]+S \right),
\end{eqnarray}
where $F_{\tilde{A}^\prime}(x,y)$ is defined for $\tilde{A}^\prime$ similarly to $F_{A^\prime}(x,y)$ in Eq.~(\ref{Pinf_Aprime}).

\section{RESULTS}

To test the analytical predictions above we conduct numerical calculations of analytic expressions,
and we compare the results with the simulation results with random regular node removal in fully interdependent networks
$\tilde{A}$ and $\tilde{B}$ with inter-similarity. All the simulation results are obtained for networks of $\tilde{N}=10^4$.

We first assume that all $A$, $B$ and $C$ are ER networks with Poissonian degree distributions.
Thus the corresponding generating functions for degree distributions
are $G_{A}(x)=e^{k_A (x-1)}$,
$G_{B}(x)=e^{k_B (x-1)}$ and $G_{0}(x)=e^{k_C (x-1)}$, respectively.

\begin{figure}
\center
\includegraphics[scale=0.5]{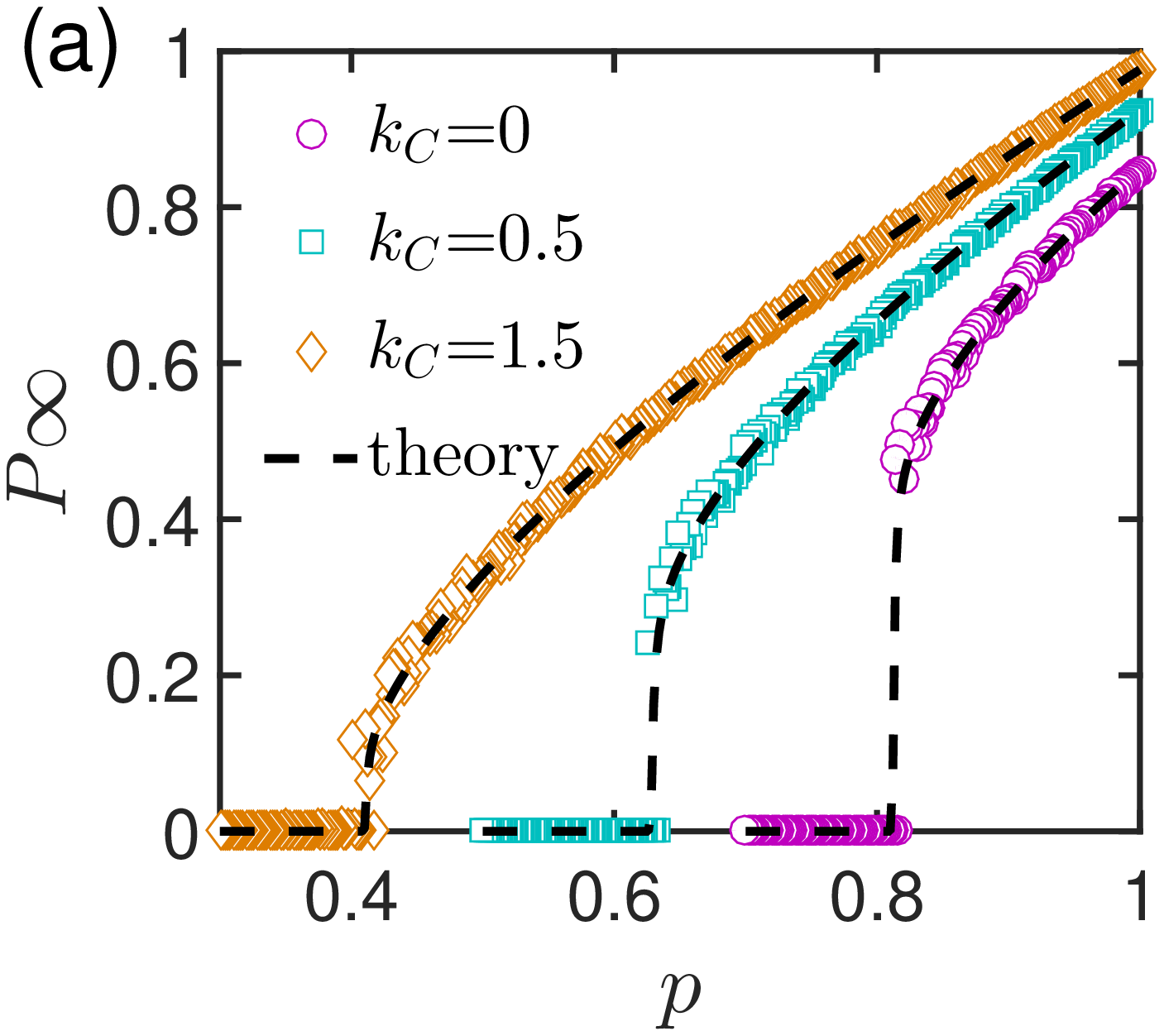}
\includegraphics[scale=0.5]{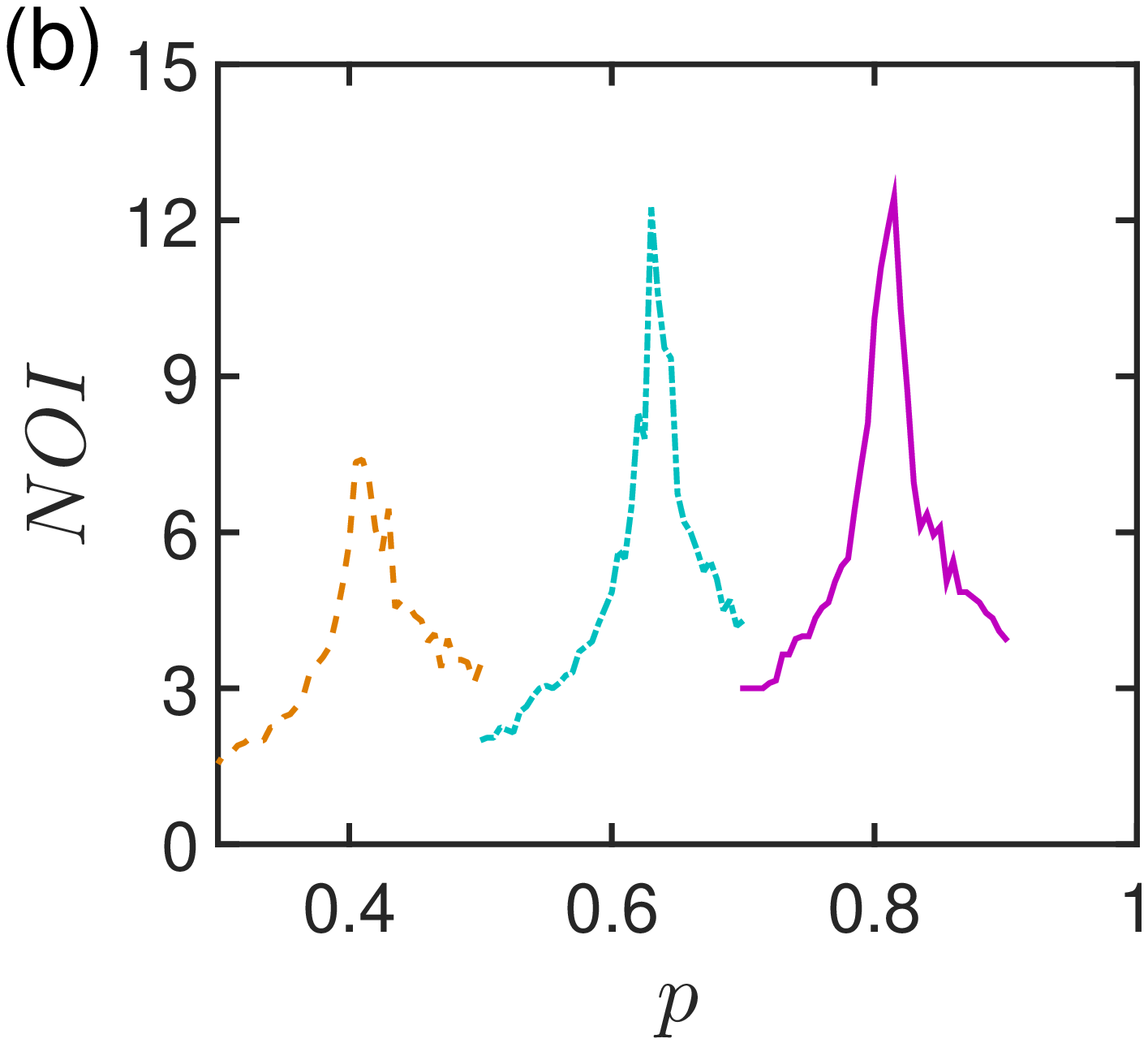}
\caption{\label{Fig6} (a) Percolation properties of networks $\tilde{A}$
and $\tilde{B}$ under random regular node removal when networks $A$, $B$ and $C$ are all ER networks. Note that simulation results (colored symbols) for $P_{\infty}^{\tilde{A}}$ (or $P_{\infty}^{\tilde{B}}$) agree well with theoretical predictions (dashed curves). Here $k_A=k_B=3$, and $k_C=0, 0.5, 1.5$ with $\tilde{N}=10^4$. (b) The number of iteration (NOI) in the simulation as a function of $p$ for each case. NOI curves peak at the critical thresholds $p_c$ at which the first-order phase transition occurs.}
\end{figure}

Fig. \ref{Fig6}(a) shows the good agreement of the numerical simulation results with the theoretical
predictions for random attacks on regular nodes with fixed $k_A=k_B=3$ and different $k_C$. Notice that for both cases $k_C<1$ and $k_C>1$, the system undergoes a discontinuous phase transition. Fig.~\ref{Fig6}(b) further shows the number of iterations (NOI) as a function of $p$. The NOI is defined as the total number of time steps in the cascading failure process. As shown in previous related works, first-order transition points correspond to peaks in the NOI curve \cite{Parshani2011,bashan2013extreme,hu2013percolation}. Therefore, Fig.~\ref{Fig6}(b) also supports the existence of first-order transitions shown in Fig.~\ref{Fig6}(a). Moreover, it is worth pointing out that as $k_C$ increases, the critical value $p_c$, at which a mutually connected giant component first arises in the system, decreases dramatically. This lends strong support to the argument that if a system of fully interdependent networks are more inter-similar, the more robust the system (with smaller transition points $p_c$) would be under random removal of regular nodes; the jump size at $p_c$ will be smaller at the meantime.

\begin{figure}
\includegraphics[scale=0.8]{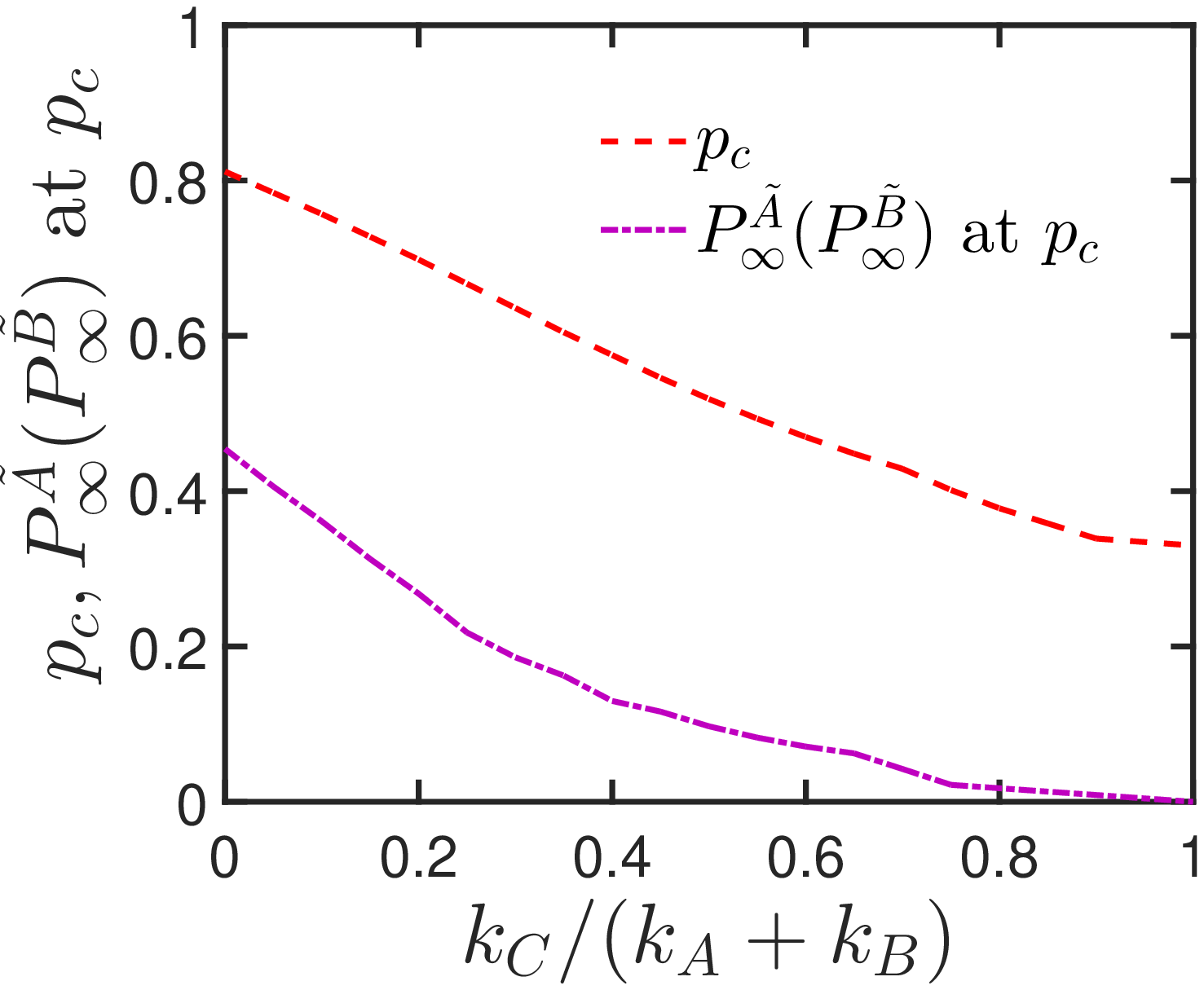}
\caption{\label{Fig7}  $p_c$ and $P_{\infty}^{\tilde{A}}$ (or $P_{\infty}^{\tilde{B}}$) at $p_c$ as a function of $\frac{k_C}{k_A+k_C}$ in theory, where $A$, $B$ and $C$ are all ER networks.
Here we increase $k_C$ from $0$ to $3$ while keeping the sum of $k_A$ and $k_C$ to be $3$. Note as
$\frac{k_C}{k_A+k_C}$ increases, $p_c$ and the jump size decrease correspondingly.
 Especially when $k_C=0$ (thus $k_A=3$) we obtain $p_c^{I}=\frac{2.4554}{k_A}$; when
 $k_C=3$ (thus $k_A=0$) we get $p_c^{II}=\frac{1}{k_C}$. These results are in agreement with previous studies \cite{min2015link,buldyrev2010catastrophic} }
\end{figure}

Moreover, to see the whole picture of the effect of inter-similarity,  we further
carried out calculations by fixing the average degree of $\tilde{A}$ and $\tilde{B}$
at 3 and increasing $k_C$ from 0 to 3. Thus, if $k_C=0$, we would just get the ordinary
interdependent network system where no inter-similarity is present; if $k_C=3$, we
get a system where all the links are common links such that networks $A_0$ and $B_0$
are exact copies of each other and thus they are essentially an ordinary
single network.  As depicted in Fig.~\ref{Fig7}, indeed, we see that if $k_C$ is close to 0,
the transition is very close to the first-order transition point of two fully interdependent
networks where pairs of interdependent nodes are randomly connected, i.e., $p_c^{I}=2.4554/k_A$. Also, as $k_C$ gets close to 3, we regain the familiar result of second-order transition point of a single network with $p_c^{II}=1/k_C$.

\begin{figure}
\center
\includegraphics[scale=0.4]{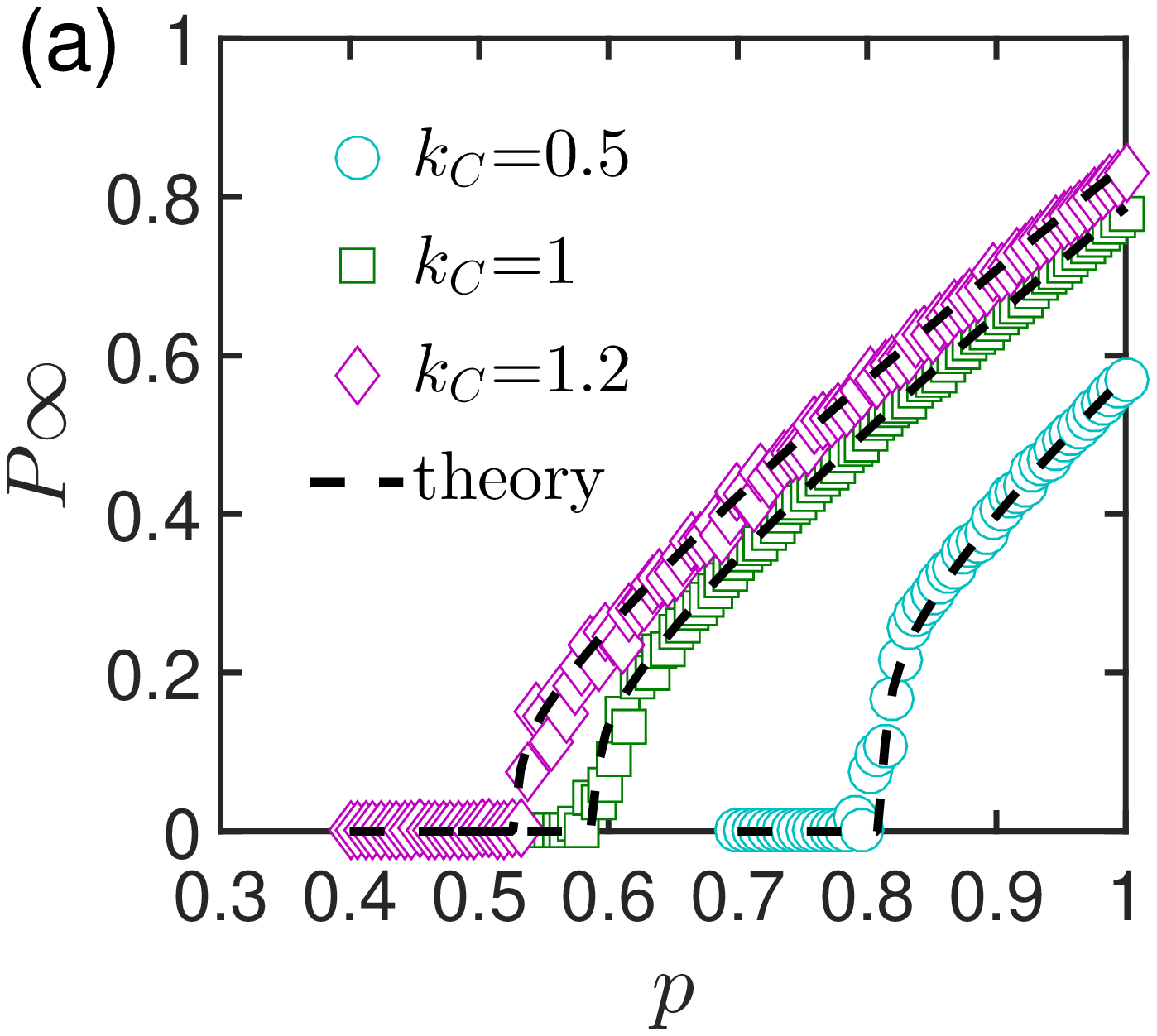}
\includegraphics[scale=0.4]{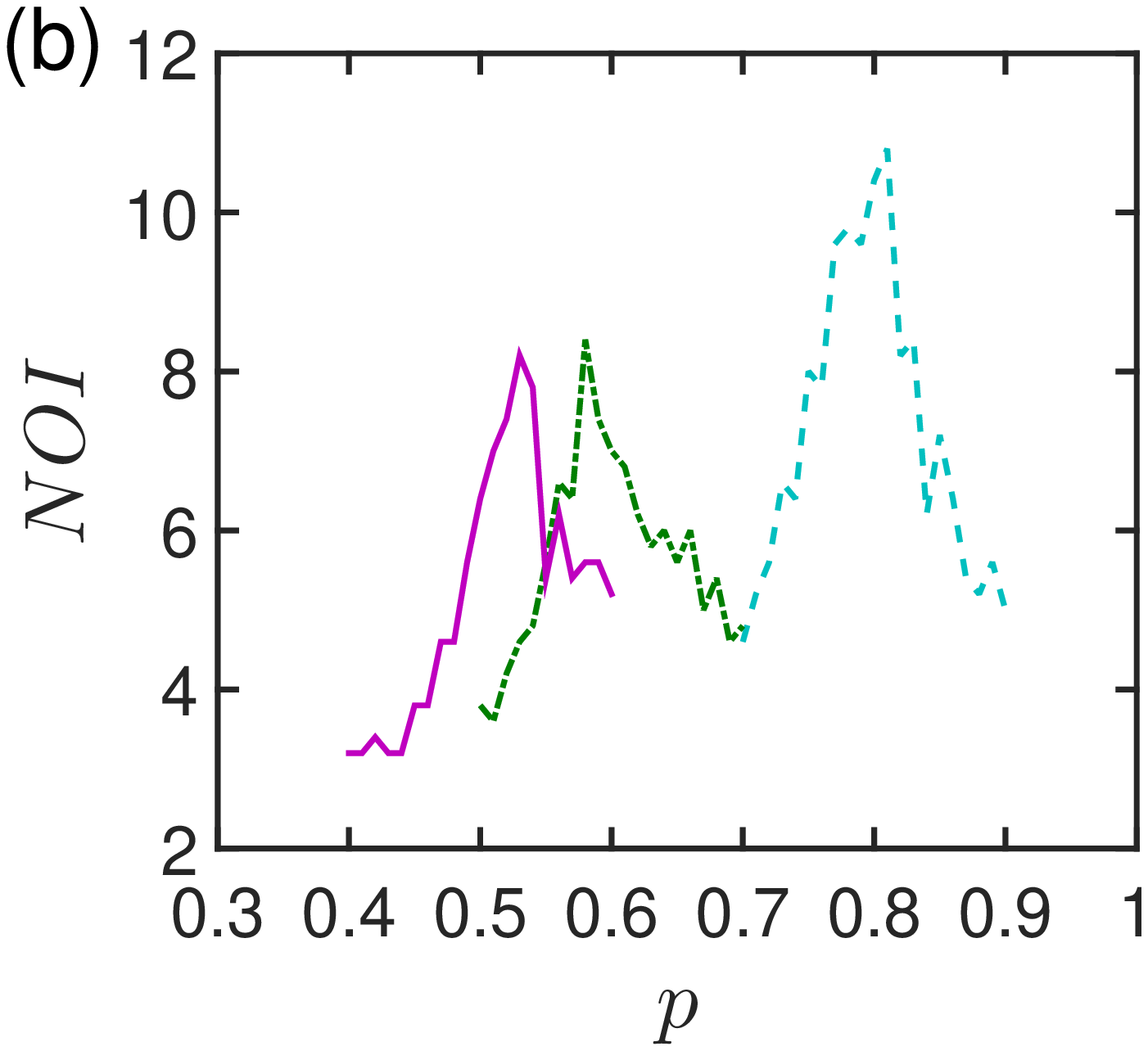}
\caption{\label{Fig8}
(a) Percolation properties of networks $\tilde{A}$ and $\tilde{B}$ where networks $A$ and $B$ are SF networks, and network $C$ is an ER network.  Note that simulation results (colored symbols)
 for $P_{\infty}^{\tilde{A}}$ (or $P_{\infty}^{\tilde{B}}$) agree well with theoretical predictions (dashed curves).
Here $k_{min}=2$, $k_{max}=50$, $\gamma=2.8$ for networks $A$ and $B$, and $k_C=0.5, 1, 1.2$ for network $C$, with system size $\tilde{N}=10^4$.
(b) The number of iteration (NOI) in the simulation as a function of $p$ for each case. NOI curves peak at the critical thresholds $p_c$ where first-order phase transitions occur.}
\end{figure}

Next similar simulations and calculations are carried out for the case where $C$ is still an ER network while $A$ and $B$ are scale-free (SF) networks with power-law degree distributions $P(k)\sim k^{-\gamma}$, with $k_{min}=2$, $k_{max}=50$ and $\gamma=2.8$. Note that numerical calculations and simulations results agree well with each other, as shown in Fig.~\ref{Fig8}. Like what we find in the case for ER networks, here for SF networks, the system always undergoes first-order phase transitions, as long as $\tilde{A}$ and $\tilde{B}$ are not built with pure common links.

\section{SUMMARY AND CONCLUSION}

To sum up, we introduce a new model of group percolation, where nodes cooperate with each other and form groups to reduce risks of failures. We have provided a general analytical approach for group percolation in fully interdependent networks by constructing the contracted networks of super-nodes. Based on this probabilistic approach, we investigate the robustness of both the original and the contracted interdependent networks. We have proved the non-existence of second order transitions for random initial attacks on either the original or the super-node networks. Moreover, we find that the system will become more robust when node group sizes increase. Finally, we have applied the proposed approach on interdependent networks with inter-similarity, where we show that the presence of inter-similarity will enhance the system robustness. Our theoretical approach sheds some new light into the percolation process of interdependent complex networks from the functionality of groups of nodes.

\section*{Acknowledgments}
We are supported by the NSFC Grant No. 61773412, Hundred-Talent Program of the Sun Yat-sen University and the Chinese Fundamental Research Funds for the Central Universities Grant 16lgjc84.

\bibliographystyle{apsrev}
\bibliography{References}

\begin{thebibliography}{33}
\expandafter\ifx\csname natexlab\endcsname\relax\def\natexlab#1{#1}\fi
\expandafter\ifx\csname bibnamefont\endcsname\relax
  \def\bibnamefont#1{#1}\fi
\expandafter\ifx\csname bibfnamefont\endcsname\relax
  \def\bibfnamefont#1{#1}\fi
\expandafter\ifx\csname citenamefont\endcsname\relax
  \def\citenamefont#1{#1}\fi
\expandafter\ifx\csname url\endcsname\relax
  \def\url#1{\texttt{#1}}\fi
\expandafter\ifx\csname urlprefix\endcsname\relax\def\urlprefix{URL }\fi
\providecommand{\bibinfo}[2]{#2}
\providecommand{\eprint}[2][]{\url{#2}}

\bibitem[{\citenamefont{Cohen and Havlin}(2010)}]{cohen2010complex}
\bibinfo{author}{\bibfnamefont{R.}~\bibnamefont{Cohen}} \bibnamefont{and}
  \bibinfo{author}{\bibfnamefont{S.}~\bibnamefont{Havlin}},
  \emph{\bibinfo{title}{Complex networks: structure, robustness and function}}
  (\bibinfo{publisher}{Cambridge University Press}, \bibinfo{year}{2010}).

\bibitem[{\citenamefont{Newman}(2010)}]{newman2010networks}
\bibinfo{author}{\bibfnamefont{M.}~\bibnamefont{Newman}},
  \emph{\bibinfo{title}{Networks: an introduction}} (\bibinfo{publisher}{Oxford
  University Press}, \bibinfo{year}{2010}).

\bibitem[{\citenamefont{Callaway et~al.}(2000)\citenamefont{Callaway, Newman,
  Strogatz, and Watts}}]{NewmanPRL2000}
\bibinfo{author}{\bibfnamefont{D.~S.} \bibnamefont{Callaway}},
  \bibinfo{author}{\bibfnamefont{M.~E.~J.} \bibnamefont{Newman}},
  \bibinfo{author}{\bibfnamefont{S.~H.} \bibnamefont{Strogatz}},
  \bibnamefont{and} \bibinfo{author}{\bibfnamefont{D.~J.} \bibnamefont{Watts}},
  \bibinfo{journal}{Phys. Rev. Lett.} \textbf{\bibinfo{volume}{85}},
  \bibinfo{pages}{5468} (\bibinfo{year}{2000}).

\bibitem[{\citenamefont{Newman et~al.}(2001)\citenamefont{Newman, Strogatz, and
  Watts}}]{NewmanPRE2001}
\bibinfo{author}{\bibfnamefont{M.~E.~J.} \bibnamefont{Newman}},
  \bibinfo{author}{\bibfnamefont{S.~H.} \bibnamefont{Strogatz}},
  \bibnamefont{and} \bibinfo{author}{\bibfnamefont{D.~J.} \bibnamefont{Watts}},
  \bibinfo{journal}{Phys. Rev. E} \textbf{\bibinfo{volume}{64}},
  \bibinfo{pages}{026118} (\bibinfo{year}{2001}).

\bibitem[{\citenamefont{Cohen et~al.}(2000)\citenamefont{Cohen, Erez, ben
  Avraham, and Havlin}}]{Shlomo2000}
\bibinfo{author}{\bibfnamefont{R.}~\bibnamefont{Cohen}},
  \bibinfo{author}{\bibfnamefont{K.}~\bibnamefont{Erez}},
  \bibinfo{author}{\bibfnamefont{D.}~\bibnamefont{ben Avraham}},
  \bibnamefont{and} \bibinfo{author}{\bibfnamefont{S.}~\bibnamefont{Havlin}},
  \bibinfo{journal}{Phys. Rev. Lett.} \textbf{\bibinfo{volume}{85}},
  \bibinfo{pages}{4626} (\bibinfo{year}{2000}).

\bibitem[{\citenamefont{Dorogovtsev et~al.}(2001)\citenamefont{Dorogovtsev,
  Mendes, and Samukhin}}]{Mendes2001PRE}
\bibinfo{author}{\bibfnamefont{S.~N.} \bibnamefont{Dorogovtsev}},
  \bibinfo{author}{\bibfnamefont{J.~F.~F.} \bibnamefont{Mendes}},
  \bibnamefont{and} \bibinfo{author}{\bibfnamefont{A.~N.}
  \bibnamefont{Samukhin}}, \bibinfo{journal}{Phys. Rev. E}
  \textbf{\bibinfo{volume}{64}}, \bibinfo{pages}{025101}
  (\bibinfo{year}{2001}).

\bibitem[{\citenamefont{Cohen et~al.}(2002)\citenamefont{Cohen, ben Avraham,
  and Havlin}}]{CohenPREpcScalefree2002}
\bibinfo{author}{\bibfnamefont{R.}~\bibnamefont{Cohen}},
  \bibinfo{author}{\bibfnamefont{D.}~\bibnamefont{ben Avraham}},
  \bibnamefont{and} \bibinfo{author}{\bibfnamefont{S.}~\bibnamefont{Havlin}},
  \bibinfo{journal}{Phys. Rev. E} \textbf{\bibinfo{volume}{66}},
  \bibinfo{pages}{036113} (\bibinfo{year}{2002}).

\bibitem[{\citenamefont{Schwartz et~al.}(2002)\citenamefont{Schwartz, Cohen,
  ben Avraham, Barab\'asi, and Havlin}}]{CohenDirectedPRE2002}
\bibinfo{author}{\bibfnamefont{N.}~\bibnamefont{Schwartz}},
  \bibinfo{author}{\bibfnamefont{R.}~\bibnamefont{Cohen}},
  \bibinfo{author}{\bibfnamefont{D.}~\bibnamefont{ben Avraham}},
  \bibinfo{author}{\bibfnamefont{A.-L.} \bibnamefont{Barab\'asi}},
  \bibnamefont{and} \bibinfo{author}{\bibfnamefont{S.}~\bibnamefont{Havlin}},
  \bibinfo{journal}{Phys. Rev. E} \textbf{\bibinfo{volume}{66}},
  \bibinfo{pages}{015104} (\bibinfo{year}{2002}).

\bibitem[{\citenamefont{Schneider et~al.}(2011)\citenamefont{Schneider,
  Moreira, Andrade, Havlin, and Herrmann}}]{Schneider2011}
\bibinfo{author}{\bibfnamefont{C.~M.} \bibnamefont{Schneider}},
  \bibinfo{author}{\bibfnamefont{A.~A.} \bibnamefont{Moreira}},
  \bibinfo{author}{\bibfnamefont{J.~S.} \bibnamefont{Andrade}},
  \bibinfo{author}{\bibfnamefont{S.}~\bibnamefont{Havlin}}, \bibnamefont{and}
  \bibinfo{author}{\bibfnamefont{H.~J.} \bibnamefont{Herrmann}},
  \bibinfo{journal}{Proc. Natl. Acad. Sci. U.S.A.}
  \textbf{\bibinfo{volume}{108}}, \bibinfo{pages}{3838} (\bibinfo{year}{2011}).

\bibitem[{\citenamefont{Buldyrev et~al.}(2010)\citenamefont{Buldyrev, Parshani,
  Paul, Stanley, and Havlin}}]{buldyrev2010catastrophic}
\bibinfo{author}{\bibfnamefont{S.~V.} \bibnamefont{Buldyrev}},
  \bibinfo{author}{\bibfnamefont{R.}~\bibnamefont{Parshani}},
  \bibinfo{author}{\bibfnamefont{G.}~\bibnamefont{Paul}},
  \bibinfo{author}{\bibfnamefont{H.~E.} \bibnamefont{Stanley}},
  \bibnamefont{and} \bibinfo{author}{\bibfnamefont{S.}~\bibnamefont{Havlin}},
  \bibinfo{journal}{Nature} \textbf{\bibinfo{volume}{464}},
  \bibinfo{pages}{1025} (\bibinfo{year}{2010}).

\bibitem[{\citenamefont{Parshani et~al.}(2010)\citenamefont{Parshani, Buldyrev,
  and Havlin}}]{parshani2010interdependent}
\bibinfo{author}{\bibfnamefont{R.}~\bibnamefont{Parshani}},
  \bibinfo{author}{\bibfnamefont{S.~V.} \bibnamefont{Buldyrev}},
  \bibnamefont{and} \bibinfo{author}{\bibfnamefont{S.}~\bibnamefont{Havlin}},
  \bibinfo{journal}{Phys. Rev. Lett.} \textbf{\bibinfo{volume}{105}},
  \bibinfo{pages}{048701} (\bibinfo{year}{2010}).

\bibitem[{\citenamefont{Parshani et~al.}(2011)\citenamefont{Parshani, Buldyrev,
  and Havlin}}]{Parshani2011}
\bibinfo{author}{\bibfnamefont{R.}~\bibnamefont{Parshani}},
  \bibinfo{author}{\bibfnamefont{S.~V.} \bibnamefont{Buldyrev}},
  \bibnamefont{and} \bibinfo{author}{\bibfnamefont{S.}~\bibnamefont{Havlin}},
  \bibinfo{journal}{Proc. Natl. Acad. Sci. U.S.A.}
  \textbf{\bibinfo{volume}{108}}, \bibinfo{pages}{1007} (\bibinfo{year}{2011}).

\bibitem[{\citenamefont{Hu et~al.}(2011)\citenamefont{Hu, Ksherim, Cohen, and
  Havlin}}]{Hu2011percolation}
\bibinfo{author}{\bibfnamefont{Y.}~\bibnamefont{Hu}},
  \bibinfo{author}{\bibfnamefont{B.}~\bibnamefont{Ksherim}},
  \bibinfo{author}{\bibfnamefont{R.}~\bibnamefont{Cohen}}, \bibnamefont{and}
  \bibinfo{author}{\bibfnamefont{S.}~\bibnamefont{Havlin}},
  \bibinfo{journal}{Phys. Rev. E} \textbf{\bibinfo{volume}{84}},
  \bibinfo{pages}{066116} (\bibinfo{year}{2011}).

\bibitem[{\citenamefont{Gao et~al.}(2012)\citenamefont{Gao, Buldyrev, Stanley,
  and Havlin}}]{gao2012networks}
\bibinfo{author}{\bibfnamefont{J.}~\bibnamefont{Gao}},
  \bibinfo{author}{\bibfnamefont{S.~V.} \bibnamefont{Buldyrev}},
  \bibinfo{author}{\bibfnamefont{H.~E.} \bibnamefont{Stanley}},
  \bibnamefont{and} \bibinfo{author}{\bibfnamefont{S.}~\bibnamefont{Havlin}},
  \bibinfo{journal}{Nat. Phys.} \textbf{\bibinfo{volume}{8}},
  \bibinfo{pages}{40} (\bibinfo{year}{2012}).

\bibitem[{\citenamefont{Li et~al.}(2012)\citenamefont{Li, Bashan, Buldyrev,
  Stanley, and Havlin}}]{li2012cascading}
\bibinfo{author}{\bibfnamefont{W.}~\bibnamefont{Li}},
  \bibinfo{author}{\bibfnamefont{A.}~\bibnamefont{Bashan}},
  \bibinfo{author}{\bibfnamefont{S.~V.} \bibnamefont{Buldyrev}},
  \bibinfo{author}{\bibfnamefont{H.~E.} \bibnamefont{Stanley}},
  \bibnamefont{and} \bibinfo{author}{\bibfnamefont{S.}~\bibnamefont{Havlin}},
  \bibinfo{journal}{Phys. Rev. Lett.} \textbf{\bibinfo{volume}{108}},
  \bibinfo{pages}{228702} (\bibinfo{year}{2012}).

\bibitem[{\citenamefont{Baxter et~al.}(2012{\natexlab{a}})\citenamefont{Baxter,
  Dorogovtsev, Goltsev, and Mendes}}]{baxter2012avalanche}
\bibinfo{author}{\bibfnamefont{G.}~\bibnamefont{Baxter}},
  \bibinfo{author}{\bibfnamefont{S.}~\bibnamefont{Dorogovtsev}},
  \bibinfo{author}{\bibfnamefont{A.}~\bibnamefont{Goltsev}}, \bibnamefont{and}
  \bibinfo{author}{\bibfnamefont{J.}~\bibnamefont{Mendes}},
  \bibinfo{journal}{Phys. Rev. Lett.} \textbf{\bibinfo{volume}{109}},
  \bibinfo{pages}{248701} (\bibinfo{year}{2012}{\natexlab{a}}).

\bibitem[{\citenamefont{Hu et~al.}(2013)\citenamefont{Hu, Zhou, Zhang, Han,
  Rozenblat, and Havlin}}]{hu2013percolation}
\bibinfo{author}{\bibfnamefont{Y.}~\bibnamefont{Hu}},
  \bibinfo{author}{\bibfnamefont{D.}~\bibnamefont{Zhou}},
  \bibinfo{author}{\bibfnamefont{R.}~\bibnamefont{Zhang}},
  \bibinfo{author}{\bibfnamefont{Z.}~\bibnamefont{Han}},
  \bibinfo{author}{\bibfnamefont{C.}~\bibnamefont{Rozenblat}},
  \bibnamefont{and} \bibinfo{author}{\bibfnamefont{S.}~\bibnamefont{Havlin}},
  \bibinfo{journal}{Phys. Rev. E} \textbf{\bibinfo{volume}{88}},
  \bibinfo{pages}{052805} (\bibinfo{year}{2013}).

\bibitem[{\citenamefont{Bashan et~al.}(2013)\citenamefont{Bashan, Berezin,
  Buldyrev, and Havlin}}]{bashan2013extreme}
\bibinfo{author}{\bibfnamefont{A.}~\bibnamefont{Bashan}},
  \bibinfo{author}{\bibfnamefont{Y.}~\bibnamefont{Berezin}},
  \bibinfo{author}{\bibfnamefont{S.~V.} \bibnamefont{Buldyrev}},
  \bibnamefont{and} \bibinfo{author}{\bibfnamefont{S.}~\bibnamefont{Havlin}},
  \bibinfo{journal}{Nat. Phys.} \textbf{\bibinfo{volume}{9}},
  \bibinfo{pages}{667} (\bibinfo{year}{2013}).

\bibitem[{\citenamefont{Cellai et~al.}(2013)\citenamefont{Cellai, L\'opez,
  Zhou, Gleeson, and Bianconi}}]{cellai6359percolation}
\bibinfo{author}{\bibfnamefont{D.}~\bibnamefont{Cellai}},
  \bibinfo{author}{\bibfnamefont{E.}~\bibnamefont{L\'opez}},
  \bibinfo{author}{\bibfnamefont{J.}~\bibnamefont{Zhou}},
  \bibinfo{author}{\bibfnamefont{J.~P.} \bibnamefont{Gleeson}},
  \bibnamefont{and} \bibinfo{author}{\bibfnamefont{G.}~\bibnamefont{Bianconi}},
  \bibinfo{journal}{Phys. Rev. E} \textbf{\bibinfo{volume}{88}},
  \bibinfo{pages}{052811} (\bibinfo{year}{2013}).

\bibitem[{\citenamefont{Zhou et~al.}(2014)\citenamefont{Zhou, Bashan, Cohen,
  Berezin, Shnerb, and Havlin}}]{ZhouDongPRE2014}
\bibinfo{author}{\bibfnamefont{D.}~\bibnamefont{Zhou}},
  \bibinfo{author}{\bibfnamefont{A.}~\bibnamefont{Bashan}},
  \bibinfo{author}{\bibfnamefont{R.}~\bibnamefont{Cohen}},
  \bibinfo{author}{\bibfnamefont{Y.}~\bibnamefont{Berezin}},
  \bibinfo{author}{\bibfnamefont{N.}~\bibnamefont{Shnerb}}, \bibnamefont{and}
  \bibinfo{author}{\bibfnamefont{S.}~\bibnamefont{Havlin}},
  \bibinfo{journal}{Phys. Rev. E} \textbf{\bibinfo{volume}{90}},
  \bibinfo{pages}{012803} (\bibinfo{year}{2014}).

\bibitem[{\citenamefont{Reis et~al.}(2014)\citenamefont{Reis, Hu, Babino,
  Andrade~Jr, Canals, Sigman, and Makse}}]{YanqingNP2014}
\bibinfo{author}{\bibfnamefont{S.~D.~S.} \bibnamefont{Reis}},
  \bibinfo{author}{\bibfnamefont{Y.}~\bibnamefont{Hu}},
  \bibinfo{author}{\bibfnamefont{A.}~\bibnamefont{Babino}},
  \bibinfo{author}{\bibfnamefont{J.~S.} \bibnamefont{Andrade~Jr}},
  \bibinfo{author}{\bibfnamefont{S.}~\bibnamefont{Canals}},
  \bibinfo{author}{\bibfnamefont{M.}~\bibnamefont{Sigman}}, \bibnamefont{and}
  \bibinfo{author}{\bibfnamefont{H.~A.} \bibnamefont{Makse}},
  \bibinfo{journal}{Nat. Phys.} \textbf{\bibinfo{volume}{10}},
  \bibinfo{pages}{762} (\bibinfo{year}{2014}).

\bibitem[{\citenamefont{Bianconi and Dorogovtsev}(2014)}]{BianconiPRE2014}
\bibinfo{author}{\bibfnamefont{G.}~\bibnamefont{Bianconi}} \bibnamefont{and}
  \bibinfo{author}{\bibfnamefont{S.~N.} \bibnamefont{Dorogovtsev}},
  \bibinfo{journal}{Phys. Rev. E} \textbf{\bibinfo{volume}{89}},
  \bibinfo{pages}{062814} (\bibinfo{year}{2014}).

\bibitem[{\citenamefont{Min et~al.}(2014)\citenamefont{Min, Yi, Lee, and
  Goh}}]{GohPRECorr2014}
\bibinfo{author}{\bibfnamefont{B.}~\bibnamefont{Min}},
  \bibinfo{author}{\bibfnamefont{S.~D.} \bibnamefont{Yi}},
  \bibinfo{author}{\bibfnamefont{K.-M.} \bibnamefont{Lee}}, \bibnamefont{and}
  \bibinfo{author}{\bibfnamefont{K.-I.} \bibnamefont{Goh}},
  \bibinfo{journal}{Phys. Rev. E} \textbf{\bibinfo{volume}{89}},
  \bibinfo{pages}{042811} (\bibinfo{year}{2014}).

\bibitem[{\citenamefont{Feng et~al.}(2015)\citenamefont{Feng, Monterola, and
  Hu}}]{feng2015simplified}
\bibinfo{author}{\bibfnamefont{L.}~\bibnamefont{Feng}},
  \bibinfo{author}{\bibfnamefont{C.~P.} \bibnamefont{Monterola}},
  \bibnamefont{and} \bibinfo{author}{\bibfnamefont{Y.}~\bibnamefont{Hu}},
  \bibinfo{journal}{New Journal of Physics} \textbf{\bibinfo{volume}{17}},
  \bibinfo{pages}{063025} (\bibinfo{year}{2015}).

\bibitem[{\citenamefont{Cellai et~al.}(2016)\citenamefont{Cellai, Dorogovtsev,
  and Bianconi}}]{Cellai2016PRE}
\bibinfo{author}{\bibfnamefont{D.}~\bibnamefont{Cellai}},
  \bibinfo{author}{\bibfnamefont{S.~N.} \bibnamefont{Dorogovtsev}},
  \bibnamefont{and} \bibinfo{author}{\bibfnamefont{G.}~\bibnamefont{Bianconi}},
  \bibinfo{journal}{Phys. Rev. E} \textbf{\bibinfo{volume}{94}},
  \bibinfo{pages}{032301} (\bibinfo{year}{2016}).

\bibitem[{\citenamefont{Baxter et~al.}(2016)\citenamefont{Baxter, Bianconi,
  da~Costa, Dorogovtsev, and Mendes}}]{Baxter2016PRE}
\bibinfo{author}{\bibfnamefont{G.~J.} \bibnamefont{Baxter}},
  \bibinfo{author}{\bibfnamefont{G.}~\bibnamefont{Bianconi}},
  \bibinfo{author}{\bibfnamefont{R.~A.} \bibnamefont{da~Costa}},
  \bibinfo{author}{\bibfnamefont{S.~N.} \bibnamefont{Dorogovtsev}},
  \bibnamefont{and} \bibinfo{author}{\bibfnamefont{J.~F.}
  \bibnamefont{Mendes}}, \bibinfo{journal}{Phys. Rev. E}
  \textbf{\bibinfo{volume}{94}}, \bibinfo{pages}{012303}
  (\bibinfo{year}{2016}).

\bibitem[{\citenamefont{Hackett et~al.}(2016)\citenamefont{Hackett, Cellai,
  G\'omez, Arenas, and Gleeson}}]{Hackett2016PRX}
\bibinfo{author}{\bibfnamefont{A.}~\bibnamefont{Hackett}},
  \bibinfo{author}{\bibfnamefont{D.}~\bibnamefont{Cellai}},
  \bibinfo{author}{\bibfnamefont{S.}~\bibnamefont{G\'omez}},
  \bibinfo{author}{\bibfnamefont{A.}~\bibnamefont{Arenas}}, \bibnamefont{and}
  \bibinfo{author}{\bibfnamefont{J.~P.} \bibnamefont{Gleeson}},
  \bibinfo{journal}{Phys. Rev. X} \textbf{\bibinfo{volume}{6}},
  \bibinfo{pages}{021002} (\bibinfo{year}{2016}).

\bibitem[{\citenamefont{Kleineberg et~al.}(2017)\citenamefont{Kleineberg,
  Buzna, Papadopoulos, Bogu\~n\'a, and Serrano}}]{Kleineberg2017PRL}
\bibinfo{author}{\bibfnamefont{K.-K.} \bibnamefont{Kleineberg}},
  \bibinfo{author}{\bibfnamefont{L.}~\bibnamefont{Buzna}},
  \bibinfo{author}{\bibfnamefont{F.}~\bibnamefont{Papadopoulos}},
  \bibinfo{author}{\bibfnamefont{M.}~\bibnamefont{Bogu\~n\'a}},
  \bibnamefont{and} \bibinfo{author}{\bibfnamefont{M.~A.}
  \bibnamefont{Serrano}}, \bibinfo{journal}{Phys. Rev. Lett.}
  \textbf{\bibinfo{volume}{118}}, \bibinfo{pages}{218301}
  (\bibinfo{year}{2017}).

\bibitem[{\citenamefont{Yuan et~al.}(2017)\citenamefont{Yuan, Hu, Stanley, and
  Havlin}}]{Yuan2017PNAS}
\bibinfo{author}{\bibfnamefont{X.}~\bibnamefont{Yuan}},
  \bibinfo{author}{\bibfnamefont{Y.}~\bibnamefont{Hu}},
  \bibinfo{author}{\bibfnamefont{H.~E.} \bibnamefont{Stanley}},
  \bibnamefont{and} \bibinfo{author}{\bibfnamefont{S.}~\bibnamefont{Havlin}},
  \bibinfo{journal}{Proc. Natl. Acad. Sci. U.S.A.}
  \textbf{\bibinfo{volume}{114}}, \bibinfo{pages}{3311} (\bibinfo{year}{2017}).

\bibitem[{\citenamefont{Liu et~al.}(2016)\citenamefont{Liu, Li, Jia, and
  Wang}}]{Liu2016}
\bibinfo{author}{\bibfnamefont{R.-R.} \bibnamefont{Liu}},
  \bibinfo{author}{\bibfnamefont{M.}~\bibnamefont{Li}},
  \bibinfo{author}{\bibfnamefont{C.-X.} \bibnamefont{Jia}}, \bibnamefont{and}
  \bibinfo{author}{\bibfnamefont{B.-H.} \bibnamefont{Wang}},
  \bibinfo{journal}{Sci. Rep.} \textbf{\bibinfo{volume}{6}},
  \bibinfo{pages}{25294} (\bibinfo{year}{2016}).

\bibitem[{\citenamefont{Baxter et~al.}(2012{\natexlab{b}})\citenamefont{Baxter,
  Dorogovtsev, Goltsev, and Mendes}}]{BaxterPRL2012}
\bibinfo{author}{\bibfnamefont{G.~J.} \bibnamefont{Baxter}},
  \bibinfo{author}{\bibfnamefont{S.~N.} \bibnamefont{Dorogovtsev}},
  \bibinfo{author}{\bibfnamefont{A.~V.} \bibnamefont{Goltsev}},
  \bibnamefont{and} \bibinfo{author}{\bibfnamefont{J.~F.~F.}
  \bibnamefont{Mendes}}, \bibinfo{journal}{Phys. Rev. Lett.}
  \textbf{\bibinfo{volume}{109}}, \bibinfo{pages}{248701}
  (\bibinfo{year}{2012}{\natexlab{b}}).

\bibitem[{\citenamefont{Newman}(2002)}]{NewmanPRE2002}
\bibinfo{author}{\bibfnamefont{M.~E.~J.} \bibnamefont{Newman}},
  \bibinfo{journal}{Phys. Rev. E} \textbf{\bibinfo{volume}{66}},
  \bibinfo{pages}{016128} (\bibinfo{year}{2002}).

\bibitem[{\citenamefont{Min et~al.}(2015)\citenamefont{Min, Lee, Lee, and
  Goh}}]{min2015link}
\bibinfo{author}{\bibfnamefont{B.}~\bibnamefont{Min}},
  \bibinfo{author}{\bibfnamefont{S.}~\bibnamefont{Lee}},
  \bibinfo{author}{\bibfnamefont{K.-M.} \bibnamefont{Lee}}, \bibnamefont{and}
  \bibinfo{author}{\bibfnamefont{K.-I.} \bibnamefont{Goh}},
  \bibinfo{journal}{Chaos, Solitons \&amp; Fractals}
  \textbf{\bibinfo{volume}{72}}, \bibinfo{pages}{49} (\bibinfo{year}{2015}).

\end{thebibliography}
\end{document}